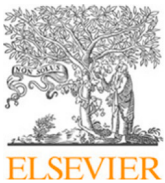

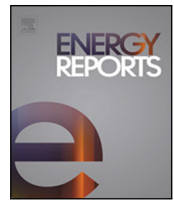

Research paper

# Cost-optimized probabilistic maintenance for condition monitoring of wind turbines with rare failures

Viktor Begun [*], Ulrich Schlickewei

*Technische Hochschule Ingolstadt, Esplanade 10, D-85049 Ingolstadt, Germany*

## A R T I C L E   I N F O

## A B S T R A C T

We propose a method, a model, and a form of presenting model results for condition monitoring of a small set of wind turbines with rare failures. The main new ingredient of the method is to sample failure thresholds according to the profit they give to an operating company. The model is a multiple linear regression with seasonal components and external regressors, representing all sensor components except for the selected one. To overcome the scarcity of the training data, we use the median sensor values from all available turbines in their healthy state. The cumulated deviation from the normal behavior model obtained for this median turbine is calibrated for each turbine at the beginning of the test period and after known failures. The proposed form of presenting results is to set a scale for possible costs, control for random maintenance, and show a whole distribution of costs depending on the free model parameters. We make a case study on an open dataset with SCADA data from multiple sensors and show that considering the influence of turbine components is more critical than seasonality. The distribution, the average, and the standard deviation of maintenance costs can be very different for similar minimal costs. Random maintenance can be more profitable than reactive maintenance and other approaches. Our predictive maintenance model outperforms random maintenance and competitors for the whole set of considered turbines, giving substantial savings.

## 1. Introduction

The worldwide use of wind energy is continuously growing. In 2000, electricity generation by wind was only about 30 TWh, accounting for a mere 0.2% of the global power share. However, by 2022, it increased to more than 2 000 TWh and a market share of 7.6% (Ember, 2023). In countries like Germany and the UK, wind power accounts for more than 20%–30% of electricity generation. This rate is even higher for smaller countries with large wind resources and installed wind power capacities, like Ireland and Uruguay – 33% in both countries and 55% in Denmark (Ember, 2023).

As this growth unfolds, there is an inevitable increase in the demand to decrease operation and maintenance costs (O&M). One can estimate the need by looking at O&M as a ratio to the levelized cost of energy (LCOE). According to Organisation for Economic Co-operation and Development, Nuclear Energy Agency (2020), the median O&M/LCOE for wind energy was 24% in 2020. It is lower than for nuclear energy worldwide, which is 38%, but still higher than for current methods extracting electricity from solar energy and gas — 13% and 11% respectively. Therefore, it is desirable to decrease O&M costs for wind turbines further. A tool for this purpose is condition monitoring (CM). By CM we mean not only detecting errors in real-time, but also predicting errors well in advance.

A choice of a CM approach largely depends on the specific task and available data. We will deal with a few highly reliable wind turbines with a rotor diameter of 90 meters and 2 MW rated power situated offshore. These conditions mean that the task is to predict a few failures well in advance and choose a cost-optimal maintenance strategy. The corresponding data are sensor values, which depend on time, are seasonal, unlabeled, partly correlated, and imbalanced. Then, a method to work with such data should be a multi-component approach containing a normal behavior model (NBM), an appropriate measure of deviation from NBM, and an alarm decision mechanism for anomaly detection.

There are many ways to build an NBM, for example, to use Kalman filter (Wei et al., 2010; Borchersen and Kinnaert, 2016), regression (Letzgus, 2020; Dao, 2021), auto-regression (Schlechtingen and Santos, 2011), support vector regression, SVR (Natili et al., 2021), use bond graphs (Djeziri et al., 2018), isolation forest (McKinnon et al., 2021), random forest (Xu et al., 2020), also gradient boosting decision tree and extreme gradient boosting, XGBOOST (Yuan et al., 2019). One can also use different neural network architectures and their combinations, like perceptron (Schlechtingen and Santos, 2011; Encalada-Dávila et al., 2021), deep belief network (Wang et al., 2019), denoising autoencoder (Du et al., 2023), a convolutional neural network, CNN, together with gated recurrent units (GRU), and with a perceptron (Kong et al., 2020), or deep autoencoder and GRU (Zhang et al., 2022), or just GRU (Encalada-Dávila et al., 2022), and so on. This variety suggests that every method used for time series prediction can also be used to make an NBM of a wind turbine.

* Corresponding author.
*E-mail addresses:* viktor.begun@gmail.com (V. Begun), ulrich.schlickewei@thi.de (U. Schlickewei).

The most popular deviation measure is the number of standard deviations above the mean. One also uses anomaly rate and more sophisticated measures like the average path length of a binary search (McKinnon et al., 2021) and Euclidian or Machalanobis distance between normal and faulty points, see Djeziri et al. (2018) and Yuan et al. (2019), Wang et al. (2019), Zhang et al. (2022). A deviation measure can be a score obtained by filtering using a kernel-based algorithm (Letzgus, 2020; Natili et al., 2021), a linear combination of features, novelty index, and principal component analysis, PCA (Natili et al., 2021). One can introduce time dependence for all the above approaches within a moving time interval and through confidence levels, for example, by using an exponential weighted moving average (EWMA) and other control charts (Dao, 2021; Xu et al., 2020; Du et al., 2023; Kong et al., 2020).

The number of possibilities for alarm mechanisms is much smaller. Most frequently, one uses some fixed threshold but also adaptive or time-dependent threshold, EWMA confidence intervals (Kong et al., 2020; Encalada-Dávila et al., 2022), or a combination of some known algorithm with machine learning, as in Zhang et al. (2022). One less popular but powerful way to encode custom decision rules is to use fuzzy logic (Schlechtingen et al., 2013; Schlechtingen and Santos, 2014). If there is enough data, any classification approach in the range from a logistic regression to a more sophisticated machine learning algorithm may work. For extensive reviews of possible CM methods, see, for example, Stetco et al. (2019), Maldonado-Correa et al. (2020), Jimenez et al. (2020), Wang et al. (2020), Black et al. (2021), Chatterjee and Dethlefs (2021), Wu et al. (2022), Badihi et al. (2022), Surucu et al. (2023).

We decided to use the Prophet (Taylor and Letham, 2018) model as a basis for our NBM because Prophet is a regression model that explicitly addresses seasonality and cross-correlations between different parts of a turbine and is already successfully used for wind turbines, see Haghshenas et al. (2023), Lyu et al. (2023), Lázaro et al. (2020). For anomaly detection, we develop our own procedure, which consists of three main steps: making a patchwork (Frankenstein[1]) turbine for training, accumulating deviations between data and forecast, and raising the alarm when it is economically reasonable for a turbine owner. The Frankenstein turbine is made as a median sensor value from available turbines in their healthy state. The accumulated deviation in our model is a cumulated sum (CUSUM) of deviations, which is calibrated (paused and then restarted with a new mean) after known failures. For different CUSUM applications, see Wei et al. (2010), Borchersen and Kinnaert (2016), Dao (2021), Xu et al. (2020), Yuan et al. (2019), Barber et al. (2022), Latiffianti et al. (2022).

Due to the scarcity of failures in our data, we do not fix an alarm threshold as in the above references but sample it with probabilities defined by the profit obtained for the taken threshold in the training data. We also propose to introduce a scale to judge the quality of obtained results between the best theoretically possible savings (maximal savings) and a case when every failure is allowed to happen and then repaired (reactive maintenance). We especially emphasize the importance of controlling for a case where turbines are inspected randomly with some frequency (random maintenance). During our research, we also noticed that most authors presented only the best results for specially optimized hyperparameters (min(cost)). It limits the possibilities of comparisons between different models and datasets. Therefore, we propose to explain and fix the hyperparameters of a model according to the taken assumptions and display the entire distribution of results as a function of other parameters. This distribution will also allow the calculation of average and standard deviation, which a model can produce.

We make a case study using an open data set (EDP - Energias de Portugal, 2018) from Energias de Portugal (EDP) to carry out a proof of concept. We found 28 works where the EDP data were analyzed, see journal publications (Barber et al., 2022; Latiffianti et al., 2022; Udo and Muhammad, 2021; Garan et al., 2022; Miele et al., 2022; Liu et al., 2021; Shahrulhisham et al., 2022; Tang et al., 2023; Jankauskas et al., 2023; Roelofs et al., 2021), conference proceedings (Du et al., 2023; Bonacina et al., 2022; Ayman et al., 2022; Pinna et al., 2022; Lauwers and De Moor, 2020; Wang and Chu, 2022; Paul et al., 2023; Pinna et al., 2023b,a; de Sá et al., 2020; Gruhl et al., 2021; Tidriri et al., 2021), a PhD (Xingchen, 2021), three master-thesis (Eriksson, 2020; Sakarvadia, 2023; Manna, 2023), and two arXiv preprints (Rabanser et al., 2022; Letgzus and Müller, 2023). However, only four of them presented the results of their algorithms in terms of savings measured in Euro (Barber et al., 2022; Latiffianti et al., 2022; Tidriri et al., 2021; Eriksson, 2020). Among them, only three report values for Hydraulic Group (Barber et al., 2022; Tidriri et al., 2021; Eriksson, 2020), which we decided to analyze.

Our approach allows us to be one of the few who have solved this challenge up to the final costs. We found that random maintenance may be a good solution in many cases, a min(cost) may be much smaller than an average cost, and we obtain the best min(cost) for Hydraulic Group among competitors for the total set of turbines. Our model allows straightforward usage for other sensor groups. A complete solution requires one more step – a decision algorithm that tells which group of sensors signals a true error if there are simultaneous signals from several groups. We briefly discuss how this can be done in summary but leave this study for future research.

The main problems that we address in this paper are the scarcity and heterogeneity of training data and the way one compares different models and presents comparison results. To overcome scarcity and heterogeneity, we propose using the Frankenstein turbine, a normal behavior model, accumulating deviations, and an adaptive threshold that is calibrated to a new turbine state after each failure. The main quality of the chosen normal behavior model is that it should not be updated too frequently, so the changes have time to build up to the values, allowing to distinguish between normal and abnormal changes in the turbine behavior and thereby detecting real failures. An important innovation is the averaging not on the level of the threshold values, but their probability distributions. Another important step is constructing the threshold distribution based on the training data and the "business" requirements on the time and costs of inspections and failures. We tried to construct the model as general as possible, so that the wind turbines mainly provide the data to the model. The purpose is to allow a generalization to other similar situations, where an expensive, reliable machine with multiple sensors produces mainly healthy data and has rare but very expensive failures that need to be predicted in time.

The paper is organized as follows: Section 2 describes the main properties of the EDP data, which determined how we constructed our model. Section 3 describes the model and all its elements. Section 4 shows the model results, while Section 5 concludes the paper. Details about CUSUM are moved to appendices.

## 2. Data

Due to the commercial sensitivity of the information contained in technical datasets, companies and firms that possess operational and failure data exhibit a profound reluctance to share such information.

---

[1] A fictional character which first appeared in Mary Shelley's novel "Frankenstein or The Modern Prometheus".

**Table 1**

Failing turbines in train, test1 and test2-period. The x2, x3 and x5 symbols mean 2, 3, and 5 failures for a same turbine.

| Group | train: year 2016 | | | | | | test1: Jan–Aug 2017 | | | | test2: Sep–Dec 2017 | | | |
|---|---|---|---|---|---|---|---|---|---|---|---|---|---|---|
| Gearbox | T01 | – | T09 | – | – | – | – | – | – | – | T06 | – | T09 | – |
| Generator | – | T06x5 | – | – | T11 | – | T07 | – | – | – | – | – | – | – |
| Generator Bearing | – | – | T07 | T09x3 | – | – | T07 | T09 | – | – | – | – | – | – |
| Hydraulic Group | – | T06 | – | – | T11 | – | T06 | T07 | – | T11 | – | T07 | T09 | T11 |
| Transformer | – | – | T07x2 | – | – | T01 | – | – | – | – | – | – | – | – |

**Table 2**

Description of failures in logs for Hydraulc Group .

| Turbine | train: year 2016 | test1: Jan–Aug 2017 | test2: Sep–Dec 2017 |
|---|---|---|---|
| T01 | – | – | – |
| T06 | Error in pitch regulation | Oil leakage in Hub | – |
| T07 | – | Oil leakage in Hub | Oil leakage in Hub |
| T09 | – | – | Pitch position error related GH |
| T11 | Hydraulic group error in the brake circuit | Hydraulic group error in the brake circuit | Hydraulic group error in the brake circuit |

However, open data are very important for developing effective algorithms and comparing their results, see discussions in Kusiak (2016), Dall-Orsoletta et al. (2022), Fernandes and Silva (2022), Jovicic et al. (2023). That is why we use the EDP data for the analysis (EDP - Energias de Portugal, 2018).

### 2.1. Description of the EDP data

EDP provides data from 5 out of 16 wind turbines with 2 MW rated power in an offshore wind park. Each turbine produces three types of data: sensors, logs, and failures. There is also sensor data from a meteorological mast (metmast). Sensors are the signals from the Supervisory Control and Data Acquisition (SCADA) system, which measures temperature, voltage, rotation speed, and similar. Metmast repeats some SCADA sensors of a turbine but gives more detailed information about the weather. It is also available when a particular turbine is not working. Logs are short text messages from a turbine. They arrive when extraordinary events happen and are timestamped with 1-second precision. They can be a source of valuable information, but they often repeat some SCADA sensors, such as Gen.ext.vent.2, temp:65 °C. Logs also come more rarely — 3 logs per turbine per hour on average, while sensor signals come with 10-minute frequency, i.e., 6 per hour. Moreover, each turbine has 83 sensors, which makes the information from sensors 83*6/3 = 166 times more abundant than from logs. Messages listed in failures by EDP have the same format as logs. Thus, the failures are the most dangerous logs. We use only sensors and failures in this paper.

EDP is interested in finding failures in 5 components: Gearbox, Generator, Generator Bearing, Hydraulic Group, and Transformer, and 16 of the 83 available sensors are associated by EDP with one of these five components. The number of sensors in each component ranges from 1 for the Hydraulic Group to 8 for the Generator. There are also 83-16=67 sensors in other components such as the Ambient, Blades, Controller, Grid, Nacelle, Production, Rotor, and Spinner. They constitute a sixth group, distinct from 5 selected components.

The data span over two full years from 01.01.2016 to 31.12.2017. We found two different time splits in the literature: one using the data from 2016 for training and the data from 2017 for testing, see Tidriri et al. (2021), Eriksson (2020), and another using 1 year and 8 months from January 1, 2016, to August 31, 2017, for training, and the remaining 4 months from September 1, 2017 to December 31,

2017, for testing, see Barber et al. (2022), Latiffianti et al. (2022). In order to compare with both found data splits, we split into 3 parts: training (train) and two test periods in 2017: a test1 ranging for 8 months from January 1 to August 31, 2017, and a test2 ranging for 4 months from September 1 to December 31, 2017. Then, the split in the Tidriri et al. (2021), Eriksson (2020) corresponds to the sum of two of our periods, test1 + test2, which we shortly call test1+2, and the split in the Barber et al. (2022), Latiffianti et al. (2022) corresponds to the test2, while we have one more testing period, test1. Having two test periods in 2017 makes even more sense for the Hydraulic Group because, with this split, the failures are distributed more homogeneously, see Tables 1 and 2. Thus, we only trained on the data from 2016 and then tested on 3 testing sets: test1, test2, and the combined test1+2.

### 2.2. Scarcity of the failure data

Available turbines are called T01, T06, T07, T09 and T11. They fail very rarely and rather chaotically. There are only 28 failures in the 5 selected components in 2 years, see Table 1, and no failures were reported outside of these components. The frequency of failures is then 28/5/2=2.8 failures per component per year. The component with the largest total number of failures is the Hydraulic Group. It has 8 out of 28 failures or 4 failures per year on average. One can also see that the failures are distributed very non-homogeneously. The Hydraulic Group failed only 2 times in 2016 (in train) but 6 times in 2017 (in test1+2). However, there is also an opposite case — the Generator failed 6 times in 2016 and only once in 2017. Looking more precisely at failures in the Hydraulic Group, one can see that there are 4 different types of these 8 failures: 'Oil leakage in Hub' (3 times), 'Hydraulic group error in the brake circuit' (3 times), and also 'Error in pitch regulation' (once) and 'Pitch position error related GH' (once).

Moreover, these failures are non-homogeneously distributed among turbines, see Table 2. A similar situation is with other components. We see the scarcity of failure data as the most important challenge in analyzing the EDP data. Then, dealing with this challenge is more important than predicting a failure in a particular component. The Hydraulic Group is the component with the largest number of failures. Therefore, we decided to study only the Hydraulic Group in this paper and develop a model that can also be applied to other components.

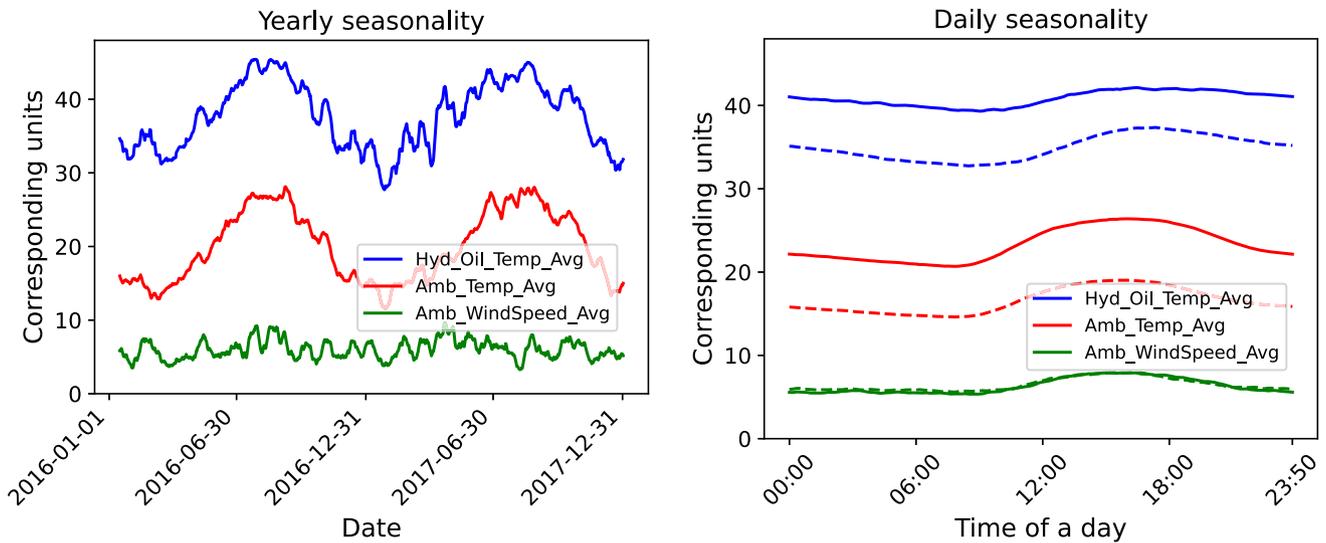

**Fig. 1.** Seasonality in the EDP data. Left: hydraulic oil temperature, outside (ambient) temperature, and wind speed (top to bottom). The units are degrees Celsius for temperatures and meters per second for the wind speed. Right: daily seasonality, the same sensors as left, but averaged over all days in "summer" (June to November, solid lines) and in "winter" (December–May, dashed lines).

## 2.3. Seasonality of the data

Another important property of the EDP data is its seasonality, see Fig. 1. In the left figure, one sees yearly seasonality. The sensor values are first averaged across all five available turbines. Subsequently, a 15-day moving average is applied to the averaged data. One can see a strong seasonality for the wind speed but rather a chaotic pattern for the oil temperature. The oil temperature resembles the outside temperature but is about 20 degrees higher and has more fluctuations. The apparent source of larger fluctuations in the oil temperature is the changing wind speed, but one can see that it is not the only source. There is a cooling and heating device in the turbine, and many other components produce heat, like the `Transformer` or cool hydraulic oil, by switching gears to lower rotation speeds in some working modes.

The Fig. 1 right shows daily seasonality. These are the same data as in Fig. 1 left, but grouped by the time of day and further averaged over "summer" and "winter". We made this split because Fig. 1 left shows only two seasons per year, one with high and another with low temperatures, and the crossing point is around June. The three selected sensor measurements show a strong and different daily seasonality. Surprisingly, there is no difference between the average wind speeds during a day in "summer" and "winter". On the contrary, the difference in the ambient temperature during the day is larger in "summer", and the difference in the oil temperature is larger in "winter".

## 2.4. EDP-cost function for hydraulic group

EDP provides its cost function, defined as a set of rules that read for the hydraulic group as follows:

$$TP_{savings} = 17\,000 \euro * N_{TP} * \frac{\Delta t}{60}, \qquad 2 < \Delta t \le 60, \tag{1}$$

$$FP_{cost} = 2\,000 \euro * N_{FP}, \tag{2}$$

$$FN_{cost} = 20\,000 \euro * N_{FN}, \tag{3}$$

$$Savings = TP_{savings} - FP_{cost} - FN_{cost}, \tag{4}$$

$$Total\,Savings = \sum_{turbine=1}^{5} Savings_{turbine}. \tag{5}$$

where $TP$, $FP$, $FN$ and $N_{TP}$, $N_{FP}$, and $N_{FN}$ are true positive, false positive, false negative, and their number correspondingly. The numbers that are written in Eqs. (1)–(3) are the costs in euros, $\Delta t$ is the integer number of days before a failure, $\Delta t \in \mathbb{Z}$. If an alarm is raised outside of the time window of $\Delta t = (2, 60]$ days, then this alarm is treated as $FP$ and costs $2\,000 \euro$.

For more clarity, we explain the physical meaning of each symbol in the cost function in more detail. True positive (TP) means finding a failure in the required time interval. It is beneficial for EDP, it is called 'savings' and is rewarded with up to $17\,000 \euro$. The number of correctly and timely identified failures is $N_{TP}$ and is less or equal to the total number of failures. A false negative (FN) means not finding a failure that actually happened. The number of false negatives $N_{FN}$ equals the number of not found failures. Finding a failure not at the right time is a false positive (FP). We treat the second and subsequent alarms within the allowed period also as false positives. Both FN and FP are disadvantageous for EDP. Such cases are called 'costs' and enter the 'savings' Eq. (4) with a negative sign. The number of false alarms, $N_{FP}$ is not formally limited, but the EDP data have 10-minute intervals, so not more than 144 alarms per day are possible, $0 \le N_{FP} \le 144$/day, as there are 24 h * 60 min/10 = 144 ten-minutes intervals per day. The price of $FP$ is 10 times smaller than that of $FN$, which means it is cheaper to have several $FP$s if they allow one not to miss a failure, the $FN$.

Let us make an example cost calculation. Imagine that there were three alarms, 61, 42, and 1 days before a failure. The alarm 42 days in advance is a TP and is rewarded by $TP_{savings} = 17\,000 \euro * 1 * 42/60 = 11\,900 \euro$. The alarms 61, and 1 days in advance are FP, so $N_{FP} = 2$, and the corresponding expenses are $FP_{cost} = 2\,000 \euro * 2 = 4\,000 \euro$. Thus, $Savings = 11\,900 \euro - 4\,000 \euro = +7\,900 \euro$. Having a correct alarm 13 instead of 42 days in advance and the same dates for other false alarms would lead to $Savings = -317 \euro$, i.e., to the losses. Not having any alarm would lead to $-20\,000 \euro$ losses.[2]

## 2.5. Data preparation and preprocessing

The main preprocessing step is that instead of 5 partly healthy turbines, we create one healthy patchwork Frankenstein turbine. It is reasonable because all the turbines are of the same type, from the same producer, presumably of the same age, and situated in the same place. Then, each healthy turbine represents a possible reaction of this type of turbine to the environmental conditions in the given place and moment. Then, the problem is finding healthy periods for each turbine. EDP rules acknowledge failure detection in a time window of 2–60 days before a failure, while a correct prediction within the last 2 days is treated as a false positive. Therefore, we cut out 58+2=60 days before each failure as unhealthy. The healthy periods are green in Fig. 2 left, while unhealthy periods are yellow and red. If several turbines work in a healthy (green) mode, we take a median. We use the median for several reasons: for simplicity, to filter outliers, because it has a clear meaning, is widely implemented, and is fast. One can also take mean values, but a median is slightly better because it corresponds to an existing turbine sensor if there is an odd number of turbines in a healthy mode. Thus, for each sensor, we take the median among the same sensors of all available turbines at each time step. Note that the obtained median sensor values in the Frankenstein turbine exhibit more fluctuations than those in the underlying turbines, as a median is selected at every time step, which is 10 min for the EDP data. Additionally, different sensors may correspond to different turbines at a particular moment.

---

[2] Later, we multiply the cost formulas by −1, so that losses obtain a positive sign and savings obtain a negative sign. With this redefinition, the cost function visually resembles a 'potential well' known in physics, and the upper cost boundary also has a positive sign, which aligns with the common perception of the word 'upper' as meaning a positive number, see Fig. 5 and discussion around it.

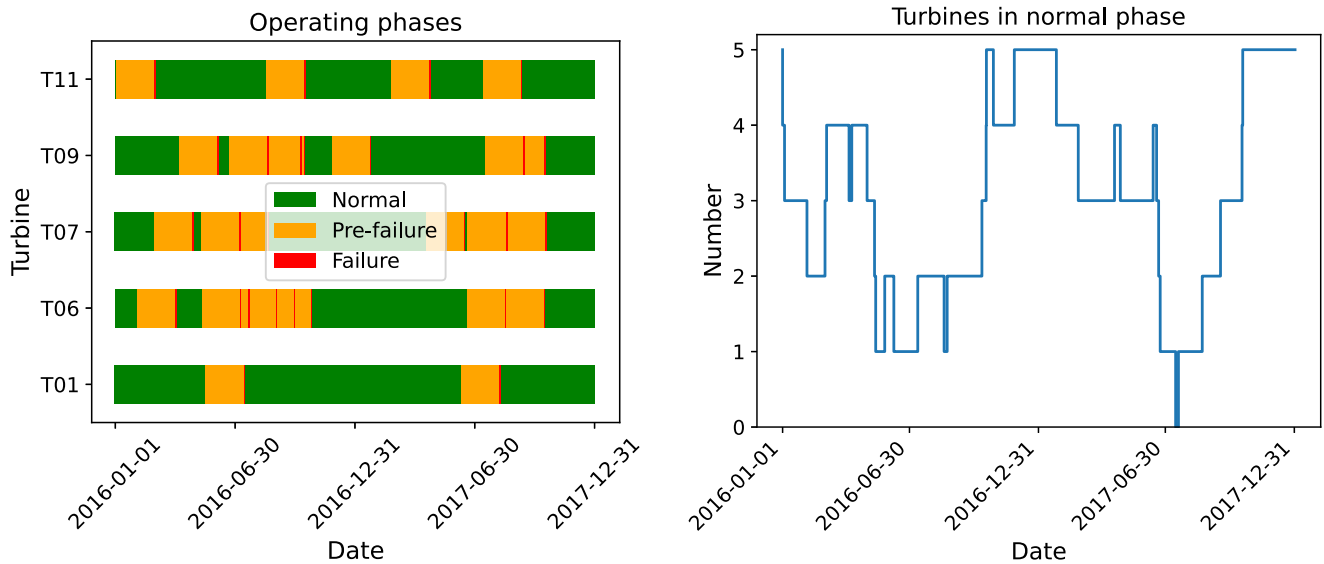

**Fig. 2.** Operating phases of the turbines (left) and the number of "healthy" turbines (right).

Thus, we assume that the cut of 60 days before failures is enough to remove the unhealthy behavior, the differences between the rest of the healthy turbines are not important, and if an unhealthy difference appears, then it will not happen for all turbines simultaneously and may be mitigated by taking the median. If the turbines were identical, then even one healthy turbine would have been enough to represent the whole wind park. The turbines are not ideal and are not in the same state. However, mixing the turbines by taking the median at every 10-minute time step decreases the dependence on a particular turbine's state. The relative position of the turbines should not be important, i.e., the wake effect should be small because the turbines are from a commercial wind park. The latter means that the position of the turbines is selected to give the maximal profit and a small loss for any wind direction. Therefore, we disregard the information about the relative position of the turbines. We also drop two dead sensors, `Prod_LatestAvg_ActPwrGen2` and `Prod_LatestAvg_ReactPwrGen2`. In order to train our algorithm to avoid seeing a false positive up to the end of the data, we add a fake "failure" at the end of each period, which has zero cost of $TP$ and $FN$ but the same price of 2 000 € for a false positive.

## 3. The model

As one may have noticed from the previous section, the EDP data set is extensive and imbalanced, sensor values are partly correlated, and there are also specific requirements for failure detection. Even after reducing the data from 5 turbines into one Frankenstein turbine, dropping logs, dead sensors, and metmast, we still have data from 81 sensors every 10 min. It makes 6 data points per hour, 144 per day, and 144*365 = 52560 points per year per sensor. This vast number of 'healthy' points contrasts with only 2–6 failures yearly. The failures are often unique and happen on different turbines with different failing histories. Logs are more abundant than failures, but they appear not for every failure, still rarely and irregularly. A turbine may experience a minor failure, labeled by a corresponding log, during harsh weather conditions in the summer of 2016 but may not be exposed to such conditions until the following summer in 2017, see Figs. 1 and 2. An essential condition is that the EDP cost function excludes failures visible less than two days in advance and favors those visible 40–60 days in advance. Such cost function means that EDP looks for failures, which build up slowly. Additionally, one has to balance the effect of finding a failure with the cost of examining a turbine.

Such a situation suggests that one needs to build a model of normal behavior, find a measure that is sensitive to small failures but robust to random, imperfect predictions, grows gradually and cumulatively when a turbine is working in a failure mode, and then find a way to raise the alarm in a right moment. Mathematically speaking, we need to find small permanent shifts in the mean, which accumulate to a value larger than a profitable threshold. Such an approach also allows to account for errors appearing suddenly if these failures do not cause immediate failures but lead to a large enough accumulated change on a scale of a few days.

### 3.1. Overview of the model

In order to solve the EDP task, we developed a multi-component approach, which combines machine learning, quality control methods, and classical probability theory. As a first step, we build a normal behavior model using Prophet (Taylor and Letham, 2018). The Prophet model can be replaced by any other machine learning model, which adequately reflects the normal behavior of a turbine and, once trained, does not change on a scale of weeks, regardless of the season. It is needed for two reasons. First, a turbine has to be exposed to all possible conditions many times so that a failure mode repeats a statistically significant number of times. Second, we accumulate the deviations as the next step of our model. We keep the original 10-minute resolution of the SCADA data, which means observing a deviation from NBM for, say, 2 weeks, then we have 2 weeks * 7 days/week * 6 points/hour/day = 84 points/hour for every hour. This accumulation of data makes a statistical analysis possible. Prophet allows such long-term predictions without a need to update the model parameters and enables a user to take into account seasonality and external regressors, i.e., the influence of other parts of a turbine, and benefit from the simplicity of the model. It also has an open-source code written and supported by the Meta Open Source community.

As the second part of our model, we sum up deviations from normal behavior and raise an alarm if an accumulated deviation is too large. For this purpose, we modified a decision interval cumulative sum (DI-CUSUM) (Hawkins and Olwell, 1998) method. The modification is that an alarm threshold is defined not as a first outlier beyond a fixed, say, 3-$\sigma$ or 6-$\sigma$ threshold, but only when it is profitable for EDP, considering the costs of inspection, replacement, and repair. Using a modified DI-CUSUM allows us to deal with the scarcity of failure data. It has only one parameter, the threshold alarm value, so fewer failures are needed to estimate the threshold statistically.

The third part of our model is needed to deal with the dependence of a DI-CUSUM method on the decisions made regarding failures. The problem is that after reporting an alarm, one has to restart the recording of the accumulated deviation with updated parameters, which may significantly influence the possibility of raising further alarms. Therefore, we sample different alarm thresholds with a probability, which is determined in the training stage based on the expected cost–benefit it gives. This sampling allows us to check all available decision paths and deliver not only the best-optimized alarm threshold but also provide the mean value, uncertainty, and a whole distribution of costs for our model. A cost distribution tells much more about a model than a single result for specially tuned parameters. It is also reasonable to expect that the cost distribution mean and uncertainty will be a much more robust way to compare different models on the same dataset.

### 3.2. Normal behavior model

The extremely small number of failures in the EDP data appears to be inadequate for direct training of any machine learning approach for classifying data points as faulty or healthy or even for statistical analysis. However, we have a lot of data in a normal state when a turbine works normally. Therefore, one can first predict normal behavior and later examine deviations from this normal behavior. We use Prophet (Taylor and Letham, 2018) as NBM, and for each sensor $i$, we make a prediction $\hat{y}_i$, which consists of four components:

$$\hat{y}_i(t) = c_i + D_i(t) + Y_i(t) + R_i(t) , \qquad (6)$$

where $c_i$ is called a *trend*, and we choose it to be a constant because we do not see a reason why the output of a sensor could rise linearly for a long time. Thus, $c_i$ means the yearly average value of a sensor $i$ in our model. The components $D_i(t)$ and $Y_i(t)$ model daily and yearly seasonality,

$$D_i(t) = \sum_{n=0}^{N_{dd}} \left( a_n \cos \left( \frac{2\pi nt}{P_{dd}} \right) + b_n \sin \left( \frac{2\pi nt}{P_{dd}} \right) \right) , \qquad (7)$$

$$Y_i(t) = \sum_{n=0}^{N_{yy}} \left( c_n \cos \left( \frac{2\pi nt}{P_{yy}} \right) + d_n \sin \left( \frac{2\pi nt}{P_{yy}} \right) \right) , \qquad (8)$$

where $N_{dd}$ and $N_{yy}$ are the numbers of daily and yearly components, and $P_{dd}$ and $P_{yy}$ are the daily and yearly periods. One can choose the parameters $N_{dd}$, $N_{yy}$, $P_{dd}$, and $P_{yy}$, while the coefficients $\{a_n, b_n, c_n, d_n\}$ are found by Prophet during the fit. The $R_i(t)$ are the external regressors:

$$R_i(t) = \sum_{j \neq i} \beta_j y_j(t) , \qquad (9)$$

where $j \neq i$ indicates that $j$ takes all values except $i$. This condition implies that we consider all sensors, except the selected one, as external influences on the chosen sensor. The coefficients $\beta_j$ show the relative importance of the selected sensor $y_i(t)$ compared to the rest of the sensors.

We fit Prophet parameters $\{a_n, b_n, c_n, d_n, \beta_j\}$ on the 2016 (train) year using Frankenstein turbine. Then we freeze the obtained parameters, select a sensor of interest $i$, take the measured $y_{j \neq i}$ values for all other sensors in the next 2017 year, and substitute them to regressors $R_i(t)$ in Eq. (9). The obtained regressors are further substituted to Eq. (6), and we obtain our model prediction $\hat{y}_i(t)$. The residual at time t for sensor i is called $\epsilon_i(t)$ and is defined as:

$$\epsilon_i(t) = y_i(t) - \hat{y}_i(t) . \qquad (10)$$

One may notice one more benefit from using Prophet with our Frankenstein turbine. Prophet is essentially a Fourier expansion and a linear regression at its core. We will use Prophet with relatively low and fixed daily and yearly seasonality for a whole train period of one year. This means we will use smooth functions like sin and cos for long periods, not shorter than a day. In contrast, sudden jumps that occur during training at junctions of the Frankenstein turbine made of healthy data appear randomly and relatively more often with intervals of as little as 10 min and are therefore smothered. However, when we use the trained model on the test data for individual turbines, we look for permanent shifts appearing on a scale larger than a day, which may be seen on the regularly oscillating background.

We do not normalize $y(t)$ to have the prediction in the original units and be able to use intuition, for example, be sure that temperatures larger than, let us say, 300 °C are not ok, or fluctuations in the size of 20 °C within an hour are not normal. This absence of normalization makes the regression coefficients $\beta_j$ dimensional and unconstrained by a usual condition for correlation to be between −1 and +1. We do not use any other functionality of Prophet, such as monthly, weekly, or other seasonality, holidays, or special events, because we do not see enough reasons to introduce it.

An example of our prediction with Prophet is shown in Fig. 3. One can see that the daily component for `daily=N_dd`=1 is a pure sine wave with a surprisingly small amplitude of about 0.9 °C. The yearly component is almost not seen in the chosen period, and its amplitude is about 0.25 °C. Domination of external regressors over seasonal components is not surprising. The external regressors already include seasonality because it is primarily reflected in the weather conditions that are depicted in the regressors. The seasonality components are there to compensate for what is not adequately represented by the external regressors. Apparently, one does not need much compensation.

We scanned the daily and yearly parameters ranging from 0 to 15 for both, including all corresponding combinations. We were looking at the mean and standard deviation (std) of the residuals for both `test1`, `test2`, and the joined `test1+2` period. We found that the number of daily components has a stronger influence on mean and std as yearly components. However, std is always larger than the difference between the obtained mean values for any combinations. This means that a choice of the daily and yearly parameters for the EDP data is somewhat arbitrary. It agrees with the finding that external regressors play the crucial role, see Fig. 3. Therefore, we select `daily=yearly=1` to address seasonality with a minimal number of parameters.

### 3.3. Deviations from normal behavior model

In order to check whether our NBM describes normal behavior well and allows us to separate it from abnormal behavior, we need to set a scale and indicators of deviation. As the indicators, we take an average of the residuals $\epsilon_{dd}(t)$ in a moving average window (MA) of $dd$-days and a standard deviation, $\sigma_{dd}$, of the residuals within the same MA. After trying different intervals, we found that visually distinct changes only occur at the scale of an order of magnitude, such as 3 and 30 days. We took 30 days for the MA of residuals because it is half of the maximal prediction interval of 60 days and compared it with the values for the first three days of the measurements. We observed that the residuals are around the starting three days average for all turbines even after two years,

$$|\epsilon_{30}(t)| < \epsilon_3 + 3 * \sigma_3 , \qquad (11)$$

where $|\epsilon_{30}(t)|$ is a modulus of the average residuals in a 30-day MA time window, while $\epsilon_3$ and $\sigma_3$ are the average residual and standard deviation in the first 3 days of training, i.e., from 1 Jan 2016 to 3 Jan 2016. This starting standard deviation $\sigma_3$ is itself around 3 °C,

$$\sigma_3 \simeq 3 \, °C , \qquad (12)$$

and mostly does not change in the following 30 days and for later MA periods for all turbines for the whole two years,

$$\sigma_3 \sim \sigma_{30}(t) , \qquad (13)$$

where $\sigma_{30}(t)$ is a standard deviation in the 30-day MA time interval. The $\sigma_{30}(t)$ almost does not decrease with a tendency to grow before some failures, being always larger than 2.1 °C and lower than 7.5 °C, which means between $0.7 * \sigma_3$ and $2.5 * \sigma_3$,

$$2.1 \, °C < \sigma_{30}(t) < 7.5 \, °C , \qquad (14)$$

$$0.7 * \sigma_3 < \sigma_{30}(t) < 2.5 * \sigma_3 . \qquad (15)$$

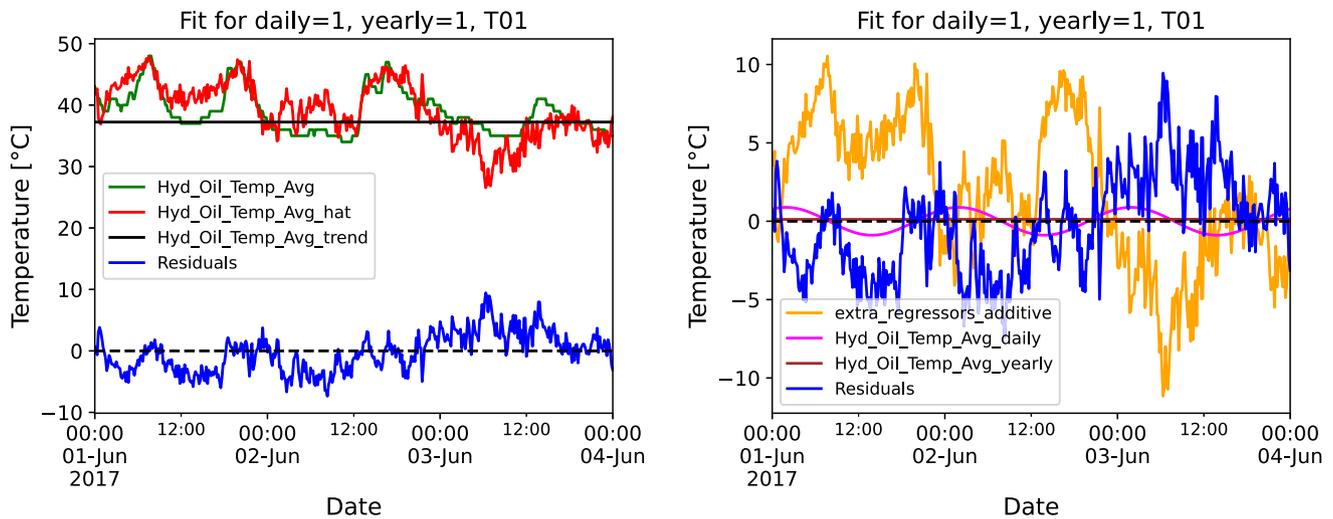

**Fig. 3.** Example of a fit with Prophet. Left, top to bottom: the measured values $y_i(t)$ where $i$ = Hyd_Oil_Temp_Avg, the average temperature of oil in the hydraulic system in the EDP notations, a forecast $\hat{y}_i(t)$, a forecast for the trend $c_i$, which is constant, and the residuals $\epsilon_i$. Right, top to bottom: the contribution of external regressors $R_i$, daily and yearly components, and the same residuals as on the left figure.

This stability of the prediction and relatively small deviations mean that NBM parameters obtained on the Frankenstein turbine in 2016 worked well for real turbines in both years, 2016 and 2017.

Limited statistics available for turbine failures in EDP data restrict our ability to use failures to find and label faulty deviations from normal behavior directly. Moreover, a more precise look at the failures shows that they may be unique because even the failures in the same category often have different error messages, and every turbine has a unique history of failures, which may be important. Fortunately, methods to address such problems have already been developed because a similar situation has arisen in the industrial production processes of goods. For example, manufacturers require a production line that can consistently deliver the desired output. They also need to halt production whenever any large enough deviation arises, regardless of the cause. Once the root cause is identified, manufacturers try to eliminate it to prevent its occurrence. This means that statistics on such failures cannot be accumulated since they are actively addressed and resolved. The corresponding discipline is statistical process control[3] (SPC), see, e.g. Montgomery (2019). A general framework of SPC is to monitor a deviation, set a tolerance (confidence level) threshold, and trigger an alarm upon crossing the threshold. The simplest way to set a confidence level in SPC is to use the number of standard deviations from an expectation. However, it is not directly applicable because one needs to adjust the alarm threshold to the EDP data and requirements for failure detection.

### 3.3.1. Calibrated CUSUM

A fundamental technique in SPC is the analysis of a cumulative sum (CUSUM) of changes — residuals in our case,

$$C_n(t) = \sum_{i=1}^{n} \epsilon(t_i) , \qquad (16)$$

where $n$ is the number of time steps after starting counting deviations. A plot of $C_n(t)$ with control limits (thresholds) is called a cumulative sum control chart. An alarm is generated when $C_n$ reaches a control limit. One can see from Eqs. (11), (15) that a popular control limit $\pm 3 * \sigma$ would not be reached for our NBM. Therefore, we define a proper threshold later in Section 3.4.

Two further important steps that we take are restarting and calibrating the CUSUM after known failures because the state of a turbine changes after failure. Calibration means that we wait for 2 days, until a new behavior of a turbine is established, record the residuals for the subsequent 2 days, calculate a mean for these two latest days, and use it as a starting point, i.e., the zero ground counting level, $\mu_0$, for a CUSUM,

$$C_n(t) = \sum_{i=1}^{n} (\epsilon(t_i) - \mu_0) . \qquad (17)$$

Thus, the calibration takes 4 days in total. After that time, a CUSUM is started, and an alarm can be triggered. We checked the behavior of $C_n(t)$ for several possible calibration time splits and found that two days of waiting plus two days for determining the ground level, $\mu_0$, is the best for the considered dataset, while the MA interval can be still 30

---

[3] There is also a similar term – statistical quality control (SQC). The SQC has quality in the name, so it is about the output, while SPC is about a process and an input. Both SQC and SPC are often used interchangeably, and there is also no difference in our context because we use the methods that are the same for both SQC and SPC.

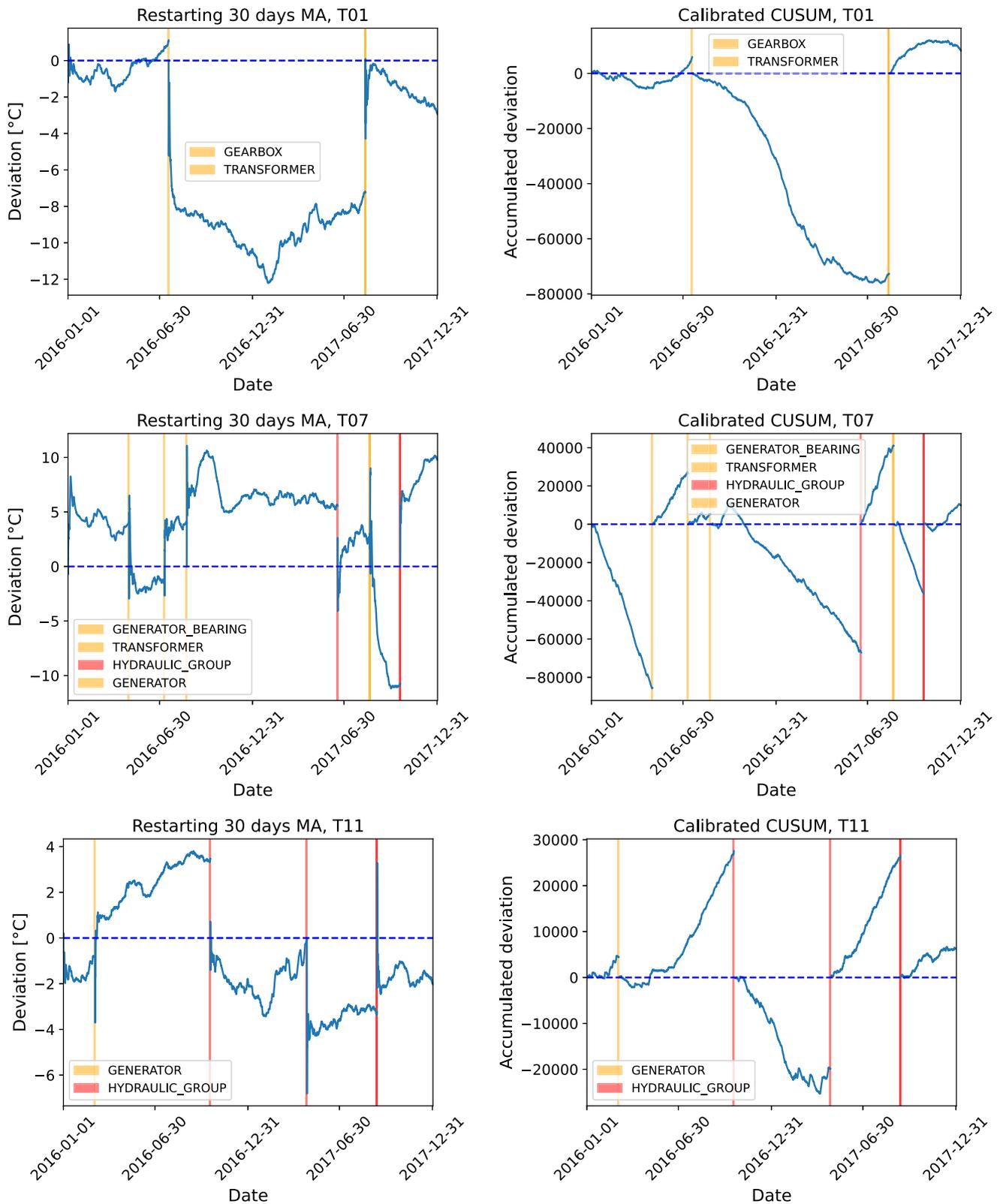

**Fig. 4.** Left: 30-day moving average of residuals for T01 and T11 turbines, which restarts after failures. Right: The corresponding restarting CUSUM, which is additionally calibrated for 2+2 days, see text. Vertical lines indicate failures. We take the same color for all non-Hydraulic Group failures because we look only at the Hydraulic Group in this paper, and other failures are not supposed to be detected by our approach.

days as in our considerations at the beginning of Section 3.3. The overall effect of restarting and calibration can be seen in Fig. 4. The T01 and T11 turbines are selected because they have zero and a maximal number of failures in the Hydraulic Group, correspondingly, while

T07 is selected for later comparison in Section 4. The restarting MA of the residuals on the left shows the same as restarting and calibrating CUSUM on the right, but calibration filters and lowers the signal while CUSUM amplifies it. For example, the visually seen persistent drop for about 10 °C after the first failure in T01 lasts for about 1 year. Without calibration, it would make a CUSUM of 10 °C/measure * 365 days * 144 measures/day = 525 600. This is an order of magnitude larger than the value reached by a calibrated CUSUM for T01 in Fig. 4. The accumulated CUSUM deviation has a formal measure of degrees Celsius or Kelvin because we did not standardize our residuals by dividing on standard deviation, not to introduce a dependence on it, because $\sigma$ is not constant. However, one may confuse the obtained CUSUM numbers with temperature. Therefore, we imply but omit writing the degree measure for CUSUMs. A similar calculation shows that the calibrated CUSUM has accumulated a drop in the mean temperature by 80 000/365 days/144 measures/day $\simeq$ 1.5 °C/measure. Thus, the calibration leads to the start of the recording of the CUSUM when the mean temperature after failure has already dropped by 8.5 °C to a new, more stable level. Therefore, we look only at the relevant shift in mean on the scale of 1.5 °C, which otherwise would have been hidden by a trivial shift for 8.5 °C caused by a previous failure. It means that the calibration is working as expected.

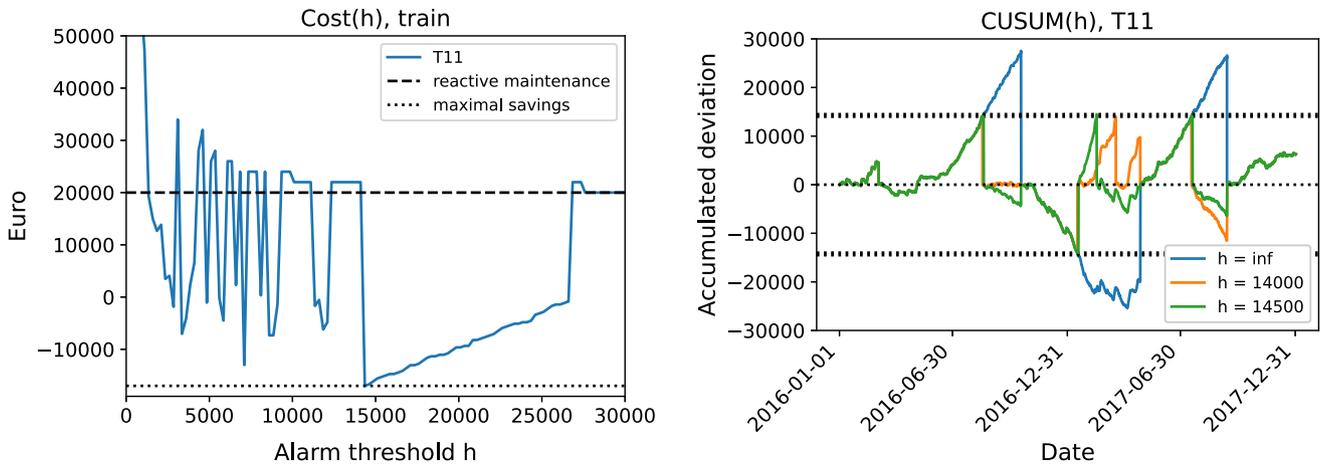

**Fig. 5.** Left: The cost of raising the alarm as a function of the height of the threshold $h$ for the T11 turbine in the `train` period. Right: CUSUM of the residuals between our model and the actual temperature values in the Hydraulic Group for the T11 turbine in the `train` period for three different alarm thresholds $h = \infty$, $h = 14\,000$, and $h = 14\,500$.

One can also see from Fig. 4 that the CUSUM for T01, which does not have failures in the `Hydraulic Group`, is larger than the CUSUM for T11, which contains failures mainly in the `Hydraulic Group`. Fortunately, false positives are cheaper than false negatives, see Section 2.4, but we still need to find a cost-effective alarm decision mechanism.

### 3.4. Alarm decision mechanism

One may notice that a CUSUM $C_n(t)$ is a distance traveled by a *random walk* with the steps $\epsilon(t_i)$. Even a very tricky random walk with sudden growths, falls, and inflection points can appear just by chance. Moreover, the size of a possible CUSUM grows with the number of steps in one direction as $C_n \sim \sigma\sqrt{n}$, where $\sigma$ is a standard deviation of the steps $\epsilon$. Even very large random CUSUMs in one direction are possible. They are less frequent but not excluded, see Appendix A. Therefore, it is crucial to correctly select a moment when we need to raise an alarm.

Due to limited statistics on turbine failures in EDP data, we need an algorithm with the lowest number of parameters, which need to be fitted. Therefore, we take the simplest version of a CUSUM, which is a two-sided CUSUM without a cutoff, and raise the alarm when it reaches a threshold of height $h$. Thus, the only parameter of this method is $h$. More sophisticated approaches can be found in Qiu (2013), Basseville and Nikiforov (1993). We make a calibration, as explained in Section 3.3.1, therefore the CUSUM has additionally a starting mean $\mu_0$ in the definition:

$$C_a = \mid \sum_{i=1}^{i_a}(\epsilon(t_i) - \mu_0) \mid \geq h \; , \tag{18}$$

where $a$ means alarm. This method has its own name - DI-CUSUM (Hawkins and Olwell, 1998), where DI stands for a **D**ecision **I**nterval, where interval means the time after the last crossing of $C_a$ with zero and subsequent growth until $h$ is reached, see Appendix B.

#### 3.4.1. Dependence of alarm decision on the path

The difference between our situation and the standard usage of a DI-CUSUM is that in the case of a statistical process control, one knows the process well and aims to raise an alarm as soon as the size of a dangerous shift is larger than a threshold. In our case, it is not possible because we need an alarm not as soon as we see a dangerous deviation but in a time window of 2–60 days before a failure. Moreover, we do not know when this 2–60 days period starts and also do not know which deviation is dangerous.

Our goal is to make a cost-effective algorithm that brings profit. Therefore, we choose the cost of selecting a threshold, $Cost(h)$, as a performance measure. This function is shown in Fig. 5 left. One can see three main regions, which are easier to explain right to left: first, the threshold is too large, the failure is not found; second, a wide range of thresholds give profit, the profit increases linearly when thresholds decrease; third, the threshold is too small, the failure may be found, but the costs depend strongly on $h$ with some random pattern. In order to understand this random pattern, let us consider two similar thresholds, $h=14\,000$ and $h=14\,500$, and plot CUSUMs for them, see Fig. 5 right. The "h=inf" line in Fig. 5 is the unaltered CUSUM, the same as in Fig. 4 for T11. The abrupt drops of the CUSUM to zero mark the moments of restarting, which happen either at failure or by reaching a threshold $h$. The form of the curves is different after restarting at different thresholds. Moreover, when we calculated the price, we saw that the $h=14\,000$ leads to 22 000 € of *losses*, while the $h=14\,500$ threshold leads to almost maximal *savings* of around 17 000 €. The reason is that $h=14\,500$ triggers an alarm at the beginning of the 60-day period, but $h=14\,000$ is reached earlier and gives an alarm earlier than 60 days. The too-early alarm gives 2 000 € of losses for the false alarm and also 20 000 € of losses for not finding the following failure because the value of the CUSUM after restart is not enough to trigger one more alarm in time.

Thus, a wide range of acceptable thresholds exist. However, one can also see from Fig. 4 that even the largest profitable threshold for T11, $h = 27\,000$, does not allow to avoid at least one false alarm for T01 because it has a much larger CUSUM. We plot cost profiles for all turbines in Fig. 6 to find the $h$-values, which are profitable for the largest number of available turbines in the `train` period. We grouped the turbines into those that failed in the `train` period (faulty) and those that did not fail in the `train` period (healthy).

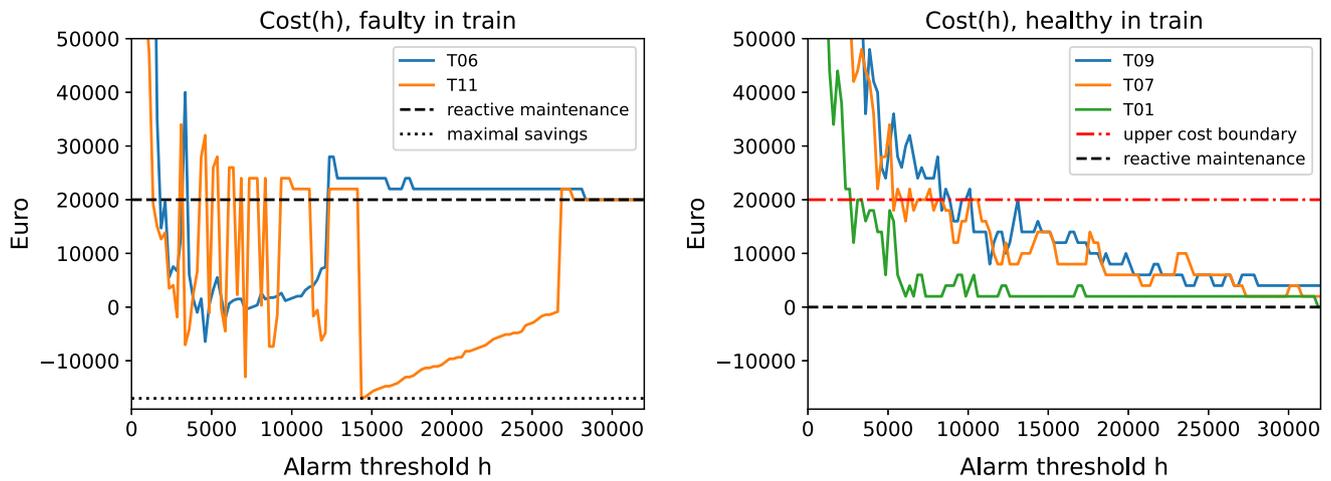

**Fig. 6.** Cost as a function of alarm threshold $h$ for faulty (left) and healthy (right) turbines in the `train` period.

One may notice from Fig. 6, left, that another faulty turbine, T06, has a similar cost profile, while cost profiles for healthy turbines look very different. If a turbine is healthy, then any alarm brings unnecessary expenses, and it would be better not to make any forecasts. In other words, `reactive maintenance` is the most profitable strategy for healthy turbines. Technically, it means having an alarm threshold $h$ so large that it is never reached. Therefore, the $Cost(h)$ for healthy turbines starts with a large number and then drops to 0€ with growing $h$. One can also see that the wide minimum in the $Cost(h)$ for T06 does not overlap with the minimum for T11. This behavior is expected because these are different turbines, and the failure `logs` are also different: `Error in pitch regulation` and `Hydraulic group error in the brake circuit`, correspondingly. The cost profiles for T07 and T09 also favor larger thresholds than for T01, and a fine for false positives disappears only after $h \gg 30\,000$ when faulty turbines give only losses. Moreover, if the turbines have different cost profiles in the `train`, then one cannot exclude large differences in other periods. It means that there is no $h$-threshold, which is the best for all turbines, and there is no guarantee that the best $h$ found in the `train` will stay the best in other periods. Therefore, we propose not to select a fixed $h$-threshold but to sample it according to our best knowledge about profitable $h$ values.

### 3.4.2. Sample thresholds

We propose to sample $h$ with a probability proportional to the savings it brings in the `train`. We accept only those $h$ that cause expected costs lower than `reactive maintenance` for a faulty turbine. This requirement can be formulated as a function of costs for each turbine "i":

$$f_i(Cost(h)) = 20\,000\ € - Cost_i(h)\ ,\qquad(19)$$

for which we accept only those $h$ that cause $f_i(Cost) > 0$. The upper cost boundary of $20\,000$€ is the price of a false negative for one faulty turbine, see Eq. (3). In shorter notations, the requirement for $h$ reads:

$$h \in \{h \mid f_i(Cost(h)) > 0\}\ .\qquad(20)$$

In order to obtain a probability distribution of acceptable thresholds, we normalize the cost function:

$$P_i(h) = \frac{f_i(Cost(h))}{\sum_h f_i(Cost(h))}\ .\qquad(21)$$

The result for faulty and healthy turbines can be seen in Fig. 7, left and right, correspondingly.

Comparing $P_i(h)$ for healthy and faulty turbines in Fig. 7, one may see that $P_i(h)$ for healthy turbines allows large $h$ and gradually prohibits small $h$, while $P(h)$ obtained from faulty turbines allows only moderate $h$. Note also that the preferred $h$ are different for the faulty turbines T06 and T11, and the overlap is relatively small. The exact form of $P_i(h)$ for healthy turbines is also different.[4] This means that the turbines fail somewhat differently, and we further need to combine the information about failures to get some average, which can be acceptably good for different turbine failures.

We have no information about the frequency of different failures in different turbines, but we know they have the same producer and characteristics. Therefore, we consider the failures as equivalent and independent and assume that a selected $h$ may work for any turbine in the future. Thus, a final probability of picking up the best $h$ is a sum of the probabilities to find the best $h$ for each turbine,

$$P(h) = \frac{\sum_i P_i(h)}{\sum_h \sum_i P_i(h)}\ ,\qquad(22)$$

where we also normalized $P(h)$ to unity, see Fig. 8. Having the distribution of thresholds $P(h)$, we can also calculate the mean and standard deviation:

$$\langle h \rangle \simeq 19\,400\ ,\qquad\text{and}\qquad \sigma_h = \sqrt{\langle h^2 \rangle - \langle h \rangle^2} \simeq 4\,400\ ,\qquad(23)$$

where $\langle h^n \rangle = \sum_h h^n P(h)$.

We can estimate the sensitivity of our model using Eq. (23). For this purpose, we divide $\langle h \rangle$ by 144 ten-minute intervals per day and by an expected permanent shift in mean, let us say by 10 °C. Then, we obtain that this shift can be seen in our model after $\langle h \rangle$°C / (144 ° intervals/day) / 10 °C ≈ 13 days on average. Similarly, one can get that the minimal permanent shift in mean, that is, on average, seen by our model 60 days after appearance, is 2.25 °C. One can also see from Fig. 8 that the distribution is relatively wide and non-Gaussian. For example, $h \sim 14\,000$ is not allowed because it was not in the original distributions for T06 and T11.

Thus, the decision model is a probabilistic generation of the alarm. It means that the alarm threshold is not fixed to some number, as in similar approaches, but the threshold value is generated from a probability distribution, and the alarm is raised once the chosen alarm threshold has been reached. The form of the distribution depends on the historical data obtained for the turbines and on the cost function imposed by the management of the wind turbine operating company. The form of the distribution that we obtained from the EDP data and the EDP cost function is shown in Fig. 8. With this distribution, one samples the thresholds. Any researcher runs an a-posteriori optimization for the best threshold parameter to describe offline data, like the data from the EDP. We argue that showing the whole distribution of the resulting costs leads to a more fair presentation of the model's saving potential. The costs resulting of 10 000 samples with the $P(h)$ distribution shown in Fig. 8 are discussed in Section 4 and are shown in Figs. 9–11.

---

[4] One may also notice a difference in scale for $P(h)$ for faulty and healthy turbines in Fig. 7. The difference appears because all distributions are normalized to 1, and the $P(h)$ distributions for healthy turbines extend long in the right direction. Then the maximum for $P(h)$ of healthy turbines is defined by the range of $h$, which we took as $h_{max} = 150\,000$ as an acceptable round number.

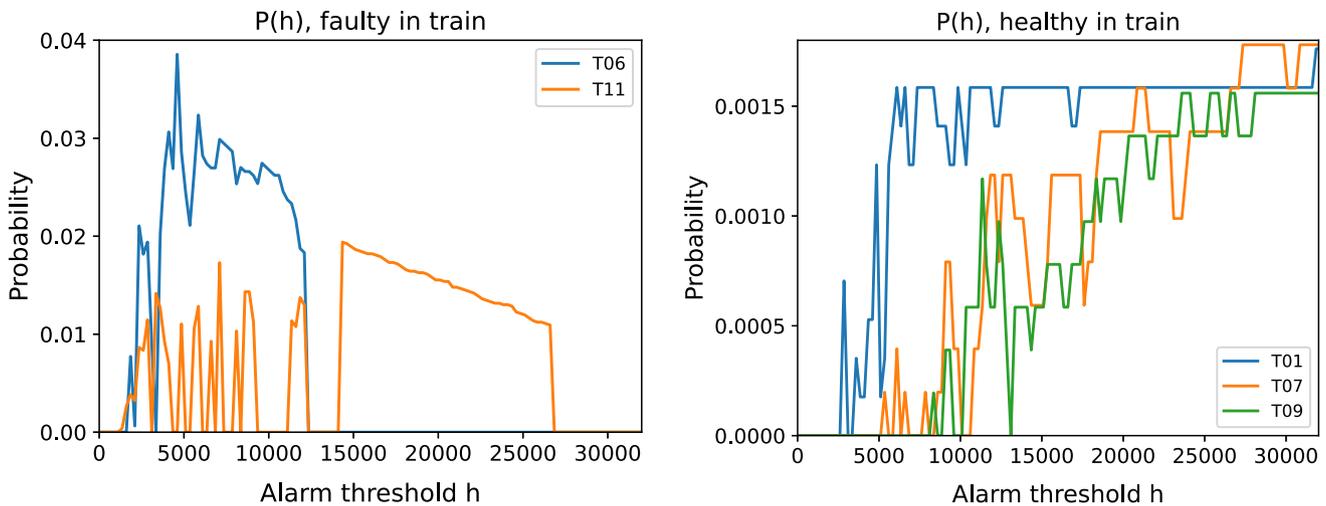

**Fig. 7.** Probability distributions to find the best price, constructed from the costs $Cost_i(h)$, see Fig. 6.

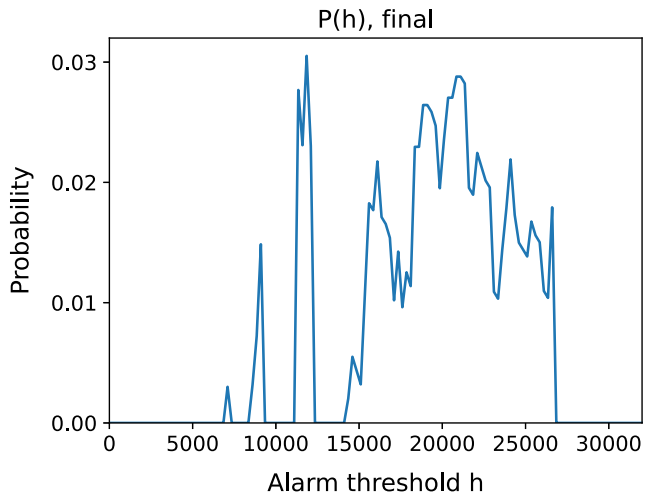

**Fig. 8.** Probability distribution to sample an alarm threshold $h$ in our model.

### 3.5. Implementation

Upon examining existing machine learning libraries with a broad scope like `scikit-learn`, as well as packages dedicated to time series prediction, see a comprehensive list and comparisons in Siebert et al. (2021), we did not find any that contain the necessary steps in the form that we need. Therefore, we developed our own Python classes and functions, which use different libraries to perform required operations. The largest code-building block is Prophet (Taylor and

Letham, 2018). We leave only daily and yearly seasonality in Prophet and use additive external regressors without upper or lower bounds on their values. We also used `NumPy`, `pandas`, `SciPy`, `Matplotlib`, `Seaborn`, and other libraries. The most computationally intensive part was finding the best seasonality parameters because their total number grows quickly with adding season resolution. It required parallelization and computing on a cluster. The rest of the computations can be done on a laptop.

## 4. Results

It is essential to establish a reference scale to discuss the results. To accomplish this, we propose considering `reactive maintenance`, `random maintenance`, and `maximal savings`, as outlined below.

### 4.1. Baseline scenarios for comparison

#### 4.1.1. Reactive maintenance

This type of maintenance is a reaction to failures that have already happened and is also called corrective maintenance. One does not find failures, $N_{TP} = 0$, $TP_{savings} = 0$, triggers no alarms and makes no inspections, $N_{FP} = 0$, $FP_{cost} = 0$, let every broken component fail, $N_{FN} = all\ failures$, $FN_{cost} = N_{FN} * 20\,000\,€$, and fix the results. This approach gives only losses, but the number of failures limits them:

$$- \ Total\ Savings \ = \ 20\,000\,€ \ * \ number\ of\ failures \ . \quad (24)$$

The number of failures in `train`, `test1`, `test2`, and `test1+2` is 2, 3, 3, and 6, correspondingly, which gives the following maximal losses after `reactive maintenance`:

$$
\begin{array}{ll}
\texttt{train}: 40\,000\,€\,, & \texttt{test1}: 60\,000\,€\,, \\
\texttt{test2}: 60\,000\,€\,, & \texttt{test1}+2: 120\,000\,€\,. \quad (25)
\end{array}
$$

#### 4.1.2. Random maintenance

In this approach, one inspects turbines randomly but with a definite distribution of failures (Gertsbakh, 2000; Nakagawa et al., 2014). To select a distribution, we first formulate our assumptions. Let us assume that inspection dates do not depend on anything, and every day has the same probability of being selected, which means the uniform distribution of inspections. Taking this distribution, we sampled 12 different dates for each turbine within two years of our data. On average, it will give one inspection per turbine per 60 days,[5] which could potentially give the best possible savings, see Eq. (1). Sampling many times from a uniform distribution, we cover even very exotic inspection patterns and see the effect of a different distribution of failures, which occurred in different periods. We also assume a 100% failure detection rate for simplicity. We show the result for 10 000 inspections in the `test2` (Sep-Dec 2017) and the `test1+2` period (the whole year 2017) in Fig. 9. Cost distributions for other periods are not shown because they are similar.

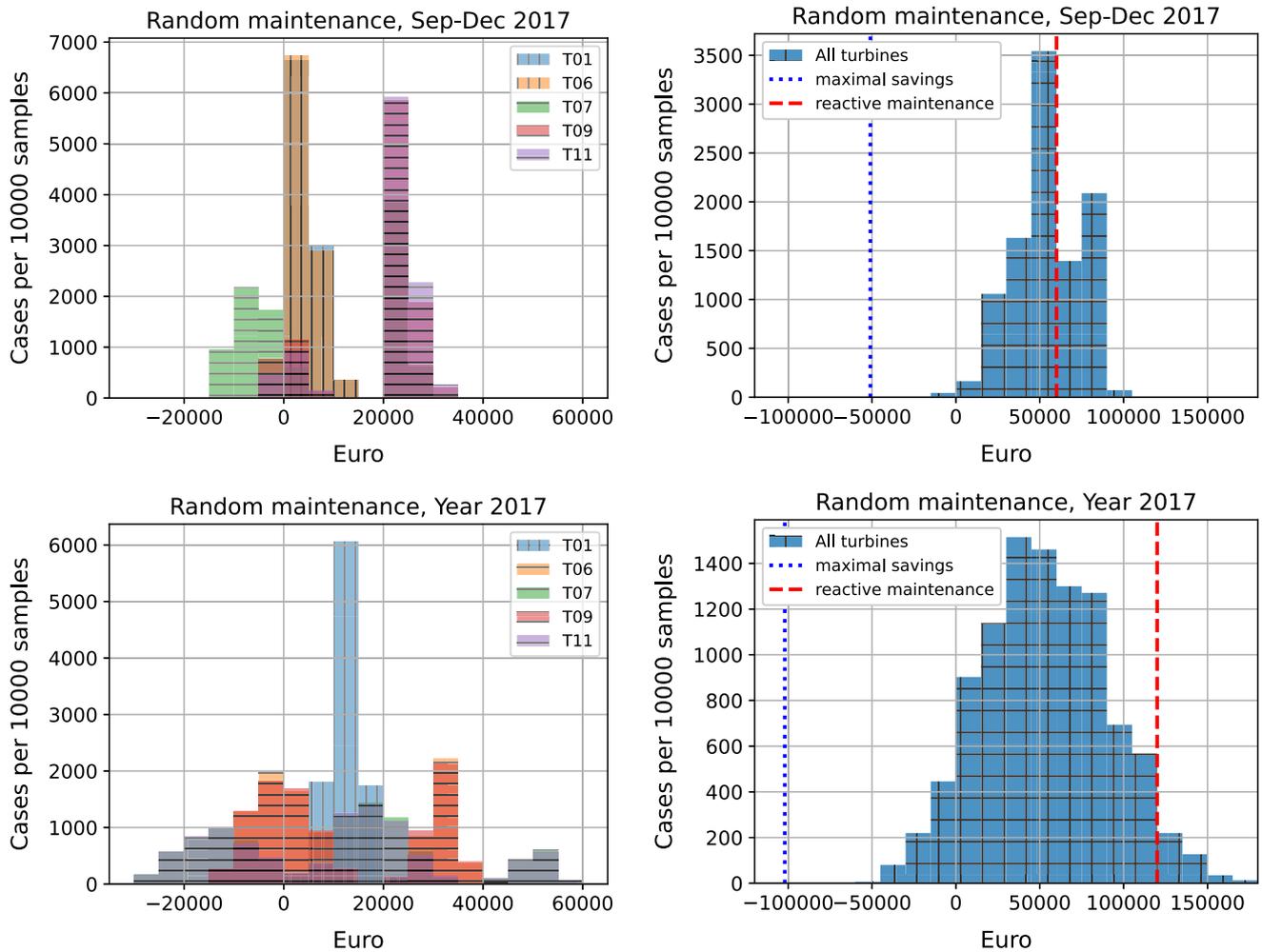

**Fig. 9.** Costs distribution of random maintenance. Left: costs for individual turbines, where healthy turbines have a vertical filling pattern, and faulty turbines have a horizontal pattern. Right: costs for the total number of turbines.

One can see that the distribution for healthy turbines has always only one maximum. The position of the mean corresponds to the number and the cost of false alarms. On average, we generated two random inspections per four months of the duration of the `test2` period and six random inspections per one year of the duration of the `test1+2` period. This sampling leads to the average costs of 4 000 € and 12 000 €, correspondingly. The distribution of costs for `random maintenance` has a width corresponding to the probability of selecting an inspection date within a chosen period on a scale of two years. It can be estimated using Binomial distribution but is larger because of additional uncertainty due to a limited number of samples. The distribution for faulty turbines has two maxima in the short `test2` period and two or three maxima in the long `test1+2` period. The reason is that there is not more than one failure per turbine in the `test2` period, but either one or two failures per turbine in the `test1+2` period. Then, there are two and three possibilities: failure found or not in `test2`, and two, one, or no failure found in `test1+2`. The costs can be estimated similarly to the `test2` period. The sum of the costs for all turbines is random and gives a smeared distribution. The widths of the distributions are relatively large. Therefore, we present the mean and standard deviation of the `random maintenance` costs rounded to a hundred euros:

$$\text{train}:\ 59\,000 + 23\,500\,, \qquad \text{test1}:\ 36\,400 + 27\,500\,,$$

$$\text{test2}:\ 54\,500 + 19\,700\,, \qquad \text{test1} + 2:\ 53\,300 + 37\,800\,, \qquad (26)$$

where the numbers are in Euro. One can see from Eq. (26) that `random maintenance` is cheaper than `reactive maintenance` in all three possible test periods, while in the `train` period, the costs of `reactive maintenance` are well covered by $1\sigma$ uncertainty. This means that savings obtained by chance can be very significant. Moreover, if one picks up only the cases that have led to the best savings for each turbine, then one can also find the cases with much lower costs among the performed 10 000 samples for the inspection dates:

$$\min(\text{train}):\ 2000\,, \qquad \min(\text{test1}):\ -37\,000\,,$$

$$\min(\text{test2}):\ -16\,100\,, \qquad \min(\text{test1} + 2):\ -54\,100\,. \qquad (27)$$

---

[5] We checked the dependence on the number of inspections and found an optimal number of about 6–10 random inspections per year for each period except for the `train`. The costs grow linearly with the increasing number of inspections in the `train` period due to the costs of false positive alarms because there are too few failures to compensate for this growth by early prediction and savings. The same effect happens in other periods for more than 10 inspections per year.

### 4.1.3. Maximal savings

With more samplings, one can push the minimal costs up to the configuration of inspection dates, which gives a theoretical minimum of zero false positives and all failures found precisely 60 days in advance:

$$\texttt{maximal savings} = -17\,000\,€ \ * \ \text{number of failures} , \quad (28)$$

which gives for all turbines in the considered periods:

$$\begin{aligned} \min_{\text{th}}(\texttt{train}) &: \ -34\,000 , & \min_{\text{th}}(\texttt{test1}) &: \ -51\,000 , \\ \min_{\text{th}}(\texttt{test2}) &: \ -51\,000 , & \min_{\text{th}}(\texttt{test1+2}) &: \ -102\,000 . \end{aligned} \quad (29)$$

One may see from Eqs. (25), (26), and (29) that the costs obtained randomly span over a huge interval starting from the maximal theoretically possible savings of $102\,000\,€$ for `random maintenance` to $120\,000\,€$ of losses for `reactive maintenance` in the `test1+2` period (the year 2017). The mean savings for `random maintenance` are notably lower than `reactive maintenance` in all test periods. Therefore, we conclude that it is not enough to present a result of a model, using the argument that it was obtained using some hyperparameters. Even if these hyperparameters are written, their choice can be somewhat random, and the result may heavily depend on how long the authors sampled from the parameter space. One should always control for `random maintenance`. The minimal check is the comparison of model results with the average `random maintenance`. Even better would also be to present a standard deviation, and the best is to show a whole distribution of costs depending on the model parameters because a cost distribution may be non-Gaussian and multi-modal.

There are too many turbines and cases to consider them separately. One is also interested in the final costs for all turbines. Therefore, we chose two turbines with interesting and somewhat opposite histories, T07 and T11, and present results for them, as well as for all turbines together, see the next Sections Section 4.2, 4.3, and 4.4.

### 4.2. A turbine with unexpected unseen failure — T07

The T07 turbine is special because it has no `Hydraulic Group` failures in the `train` but has the same failure in both `test1` and `test2` periods. Moreover, the corresponding `log` says that it is `Oil leakage in Hub`, which did not happen in any turbine in the `train` period, see Tables 1 and 2. Therefore, we can check how our model generalizes and how different period duration influences the detection. The costs for all possible test periods and available models are presented in Table 3, with average and standard deviation (when available) rounded to hundreds of euros. We show the distributions of costs in the `test2` and `test1+2` periods in Fig. 10. We also show the same for the T11 turbine to compare T07 and T11 better later in Section 4.3.

Our model gives better results than `random maintenance`, especially in the `test2` period from September to December 2017. The failure of T07 is always found in our model, while with `random maintenance`, the failure is sometimes found (the maximum below zero), but sometimes not (the maximum above $20\,000\,€$), the same as in Fig. 9 upper left for T07. The mean cost obtained in our model is lower than the lower quartile for `random maintenance`.

The costs for 2017 occasionally give almost the same average in our model and for `random maintenance`, while the distributions are different. There are two failures in T07, and `random maintenance` gives three maxima. These maxima correspond to the cases where both failures are found, one found but one not, and both not found, as shown in Fig. 9 bottom left for T07. However, our model gives only two maxima at the lowest costs, so we always find at least one of the two failures. Note that both `random maintenance` and our model are much better than `reactive maintenance`.

Only authors of Barber et al. (2022) and Tidriri et al. (2021) presented their results for T07. The authors of Tidriri et al. (2021)

considered only the `test1+2` period, while the authors of Barber et al. (2022) do not publish the costs in a table but show figures with alarms and failures as a function of time. From these figures, we find the number of false positives, $N_{FP}$, and also the approximate number of days between alarms and failures, $\Delta t$. From these two numbers, we calculate the costs obtained in their models in all possible test periods, see Table 3. These costs are their best results. Therefore, we should compare them with our best results, which give maximal savings (`min(cost)`) among all $10\,000$ samples we have taken. The models are ranged according to the `min(cost)`. We also added the corresponding numbers of days between alarms and failures and the numbers of false positives and false negatives.

One can see from Table 3 that the failure in the `test1` period is found only in our model or randomly. This failure is the `Oil leakage in Hub` unseen in the `train`. The average cost is relatively high. There are two reasons: first, a problem with `Hydraulic Group`, which is seen at the beginning of the `test1` period in Fig. 4 but was solved by an operator, which is seen from the `logs` like `User 0 primary access` and other `logs` indicating manual interventions in that period; second, there was heating in August 2017, that was caused by a failure in generator bearing on `2017-08-20 06:08:00` with a clear log `Generator bearings damaged` and the subsequent failure in the generator the next day `2017-08-21 14:47:00` with a log `Generator damaged`. The heating was apparently so intense that the hydraulic oil temperature sensor also measured it. However, the average cost is smaller than the `reactive maintenance` approach. Therefore, our model is profitable in the `test1` period.

The next failure with the same `log` appearing in the `test2` period is much better found, which can be attributed to the behavior of MA and CUSUM in Fig. 4. In the joined `test1+2` period, the authors of Barber et al. (2022) see the second failure better than we do. However, we find both failures, which makes the `min(cost)` in the joined period smaller, while our average costs are also below `reactive maintenance`. Nevertheless, `random maintenance` is also a competitive approach in this case.

If an algorithm is used in production, it faces not the best but all possible situations, producing a distribution of outcomes. Therefore, we would like to emphasize that it is very important to show not only the best results but also to present the whole distribution of outcomes and remember to check the model against `random maintenance`.

The example of turbine T07 shows that the Frankenstein turbine is indeed a solid base to train the normal behavior model. The T07 had significant downtimes during the training period and, therefore, did not contribute much training data for the normal behavior model. Yet, with our method, we could generate up to $30\,500$ € savings of the average maintenance cost for turbine T07 in 2017.

### 4.3. A turbine, which fails similarly in all periods — T11

It is interesting to look at T11 because it is the only turbine that fails in the `train`, `test1`, and `test2` periods. Moreover, the failure log is always the same: `Hydraulic group error in the brake circuit`. This repetition could have been a reason to find the failure better than for T07. Surprisingly, this is not the case, because we found only one article considering T11, and this article reports losses larger than `reactive maintenance`, see Table 4.

Our model finds failures for T11 similarly well as for T07 but a little more difficult for T11. The reason is that the corresponding CUSUM for T11 is smaller than for T07, see Fig. 4. The reason for `random maintenance` is that for T11, the failure appears only 11 days after the start of the period, while it was 48 days for T07. Then, the chances of randomly choosing a day in this period and the total possible savings are also correspondingly smaller. On average, failures in T11 are found very well by our model. The mean of each distribution is far below the `reactive maintenance` threshold, which means profitability.

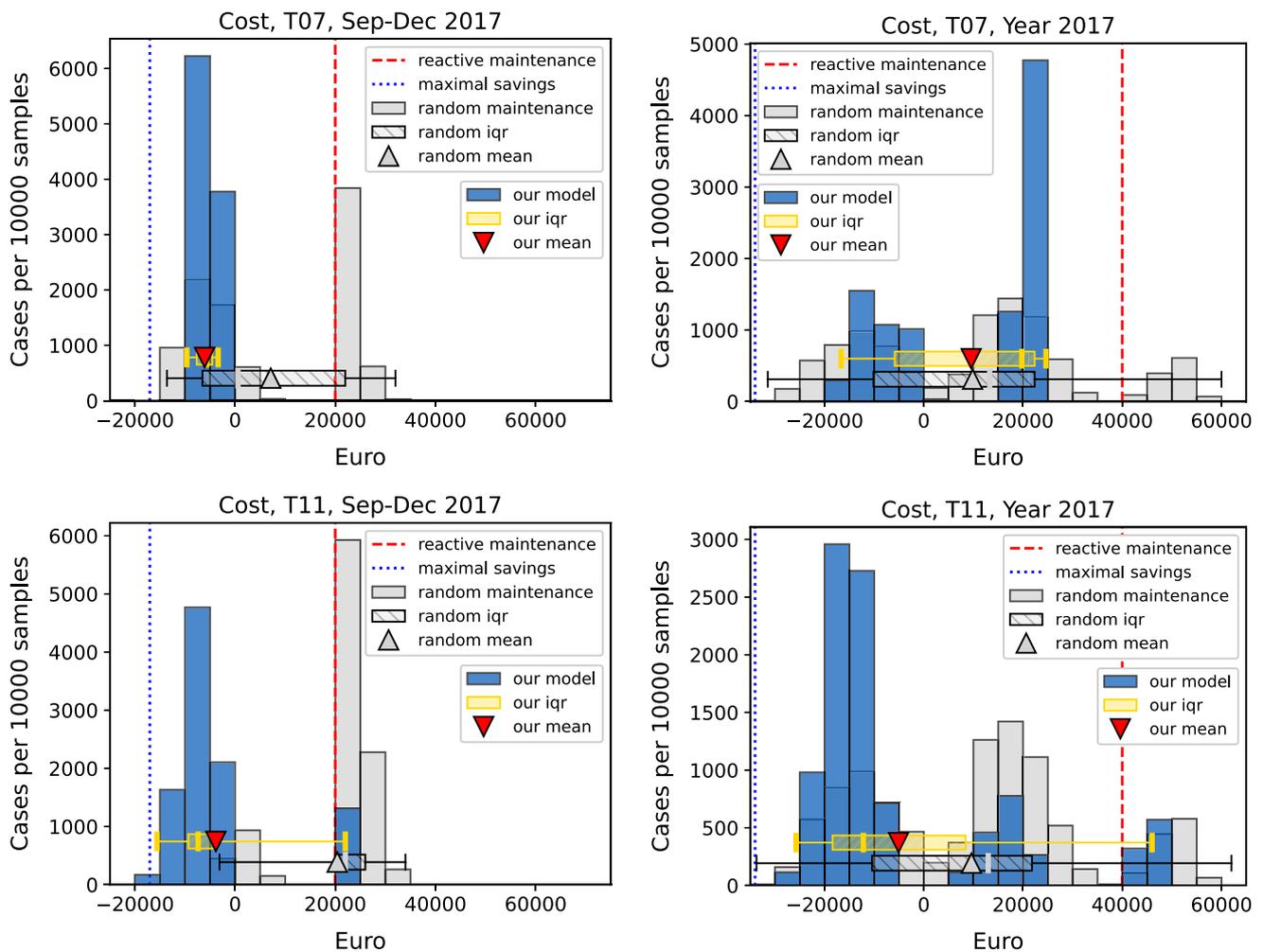

**Fig. 10.** Distribution of costs in our model vs `random maintenance` for T07 (upper row) and T11 turbines (lower row). Two periods are considered — `test2`, Sep-Dec 2017 (left), and `test1+2`, the whole 2017 year (right). Vertical lines limit the range between `maximal savings` (dotted), which are theoretically possible, see also $min_{th}$ in Eq. (29), and the costs of `reactive maintenance` (dashed). Up and down triangles show the mean values for `random maintenance` and in our model, correspondingly. Shaded rectangles and a small vertical line within them show interquartile range (iqr) and median, correspondingly. Whiskers are set as minimal and maximal values.

### 4.4. All turbines together

The sum of the results for all turbines is shown in Fig. 11 and Table 5. The sum is shuffled, which means that each turbine is inspected individually, and then the inspection plans for different turbines are combined randomly.

Both `random maintenance` and our model are profitable on average in all test periods. The best detection time per failure is more than 40 days in all test periods in our model. Reducing the number of false alarms may be very profitable for long observation periods. Interquartile ranges are relatively small, but standard deviations are large, especially whiskers extending far away from the mean. It indicates that the distributions have heavy tails. In other words, occasional large savings, as well as rare large losses, are possible. The failures are better found in our model than by `random maintenance`, with the most significant difference in September–December 2017. Our model and `random maintenance` give better results than those found in the literature and are below `reactive maintenance`, which means profitability.

Reminding the fault Table 2, one may notice that there are four types of faults, but, only one of them, the 'Hydraulic group error in the brake circuit' repeats in the `train` and in all `test` periods, while two new unseen errors appear in `test`. These errors sound and also manifest differently, while before the 'Pitch position error...' the hydraulic temperature sensor is not heating as strongly as before 'Oil leakage...'. Nevertheless, our model allows finding the unseen new faults, and the netto-cost balance is not only larger than the `reactive maintenance` but even positive. This means that our model gave profit also in the new unseen cases.

## 5. Summary and outlook

We analyzed the data from a few wind turbines with long periods of normal behavior and a few heterogeneously distributed failures (EDP - Energias de Portugal, 2018). Our model consists of four parts: data preparation, normal behavior model, quantifying deviations from normal behavior, and a probabilistic alarm decision mechanism. First, we make one healthy Frankenstein turbine for reference. We cut presumably faulty periods for each turbine and take median values of the rest for each sensor. Second, to disentangle the influence of different turbine components, we treat all sensors except those chosen as external regressors in a linear regression. Then, we use one year of Frankenstein data as input to find the regression coefficients. Using these coefficients, we predict for the whole of the following year. For this purpose, we used the Prophet model (Taylor and Letham, 2018), which can handle the mentioned steps and has a built-in daily and yearly seasonality. Third, we assume that a failure can be found as a slight persistent shift in mean between prediction and outcome and use a cumulative sum of residuals to find such shifts. Due to the scarcity of the failure data, we take DI-CUSUM (Hawkins and Olwell,

**Table 3**

Comparison of different models for T07 turbine in all test periods. Columns left to right: the model; the mean and standard deviation of the cost distribution (if available); minimal costs (`min(cost)`), which are also the maximal savings; the average number of days in advance that failures were detected, and the number of false positives and false negatives corresponding to `min(cost)`.

| Models for T07 | Cost in Euro | | $\langle \Delta t \rangle$ | $N_{FP}$ | $N_{FN}$ |
|---|---|---|---|---|---|
| | average ± std | `min(cost)` | | | |
| *test1, Jan-Aug 2017, 1 failure* | | | | | |
| maximal savings | −17 000 | −17 000 | 60 | 0 | 0 |
| random maintenance | 6700 ± 15 500 | −17 000 | 60 | 0 | 0 |
| our model | 15 700 ± 14 200 | −9000 | 60 | 4 | 0 |
| reactive maintenance | 20 000 | 20 000 | 0 | 0 | 1 |
| Barber et al. (2022): NBM, NBM-LI, WHC-LOF | – | 20 000 | 0 | 0 | 1 |
| Barber et al. (2022): CCA, EDP | – | 22 000 | 0 | 1 | 1 |
| Barber et al. (2022): LoMST | – | 24 000 | 0 | 2 | 1 |
| *test2, Sep-Dec 2017, 1 failure* | | | | | |
| maximal savings | −17 000 | −17 000 | 60 | 0 | 0 |
| Barber et al. (2022): WHC-LOF | – | −17 000 | 60 | 0 | 0 |
| random maintenance[a] | 7100 ± 14 500 | −13 600 | 48 | 0 | 0 |
| Barber et al. (2022): LoMST | – | −13 583 | 55 | 1 | 0 |
| our model | −6100 ± 1800 | −9633 | 34 | 0 | 0 |
| reactive maintenance | 20 000 | 20 000 | 0 | 0 | 1 |
| Barber et al. (2022): NBM, NBM-LI, CCA, EDP | – | 20 000 | 0 | 0 | 1 |
| *test1+2, year 2017, 2 failures* | | | | | |
| maximal savings | −34 000 | −34 000 | 60 | 0 | 0 |
| random maintenance | 9800 ± 21 600 | −31 433 | 59 | 1 | 0 |
| our model | 9500 ± 15 000 | −16 650 | 43.5 | 4 | 0 |
| Barber et al. (2022): WHC-LOF | – | 3000 | 30 | 0 | 1 |
| Barber et al. (2022): LoMST | – | 10 416 | 26 | 3 | 1 |
| reactive maintenance | 40 000 | 40 000 | 0 | 0 | 2 |
| Tidriri et al. (2021), Barber et al. (2022): NBM, NBM-LI | – | 40 000 | 0 | 0 | 2 |
| Barber et al. (2022): CCA, EDP | – | 42 000 | 0 | 1 | 2 |

[a] The failure in T07 appears 48 days after the start of the `test2` period. Therefore, the costs for random maintenance are limited by the earliest detection of 48 days and corresponding savings of 13 600 €. It is less than the earliest allowed detection in 60 days, but EDP did not mention how to count costs in this case and both Barber et al. (2022) and Tidriri et al. (2021) do not discuss this issue. Therefore, we decided to show the best results in all models. Otherwise, the failure in `test2` has to be treated as not found in both Barber et al. (2022) and Tidriri et al. (2021).

**Table 4**

Comparison of different models for T11 turbine in all test periods, see Table 3 for description of columns.

| Models for T11 | Cost in Euro | | $\langle \Delta t \rangle$ | $N_{FP}$ | $N_{FN}$ |
|---|---|---|---|---|---|
| | average ± std | `min(cost)` | | | |
| *test1, Jan-Aug 2017, 1 failure* | | | | | |
| maximal savings | −17 000 | −17 000 | 60 | 0 | 0 |
| random maintenance | 6800 ± 15 600 | −17 000 | 60 | 0 | 0 |
| our model | 400 ± 12 300 | −13 000 | 60 | 2 | 0 |
| reactive maintenance | 20 000 | 20 000 | 0 | 0 | 1 |
| *test2, Sep-Dec 2017, 1 failure* | | | | | |
| maximal savings | −17 000 | −17 000 | 60 | 0 | 0 |
| our model | −3800 ± 10 500 | −15 583 | 55 | 0 | 0 |
| random maintenance[a] | 20 400 ± 8400 | −3117 | 11 | 0 | 0 |
| reactive maintenance | 20 000 | 20 000 | 0 | 0 | 1 |
| *test1+2, year 2017, 2 failures* | | | | | |
| maximal savings | −34 000 | −34 000 | 60 | 0 | 0 |
| random maintenance | 9600 ± 21 600 | −33 716 | 59.5 | 0 | 0 |
| our model | −5100 ± 19 900 | −25 750 | 52.5 | 2 | 0 |
| reactive maintenance | 40 000 | 40 000 | 0 | 0 | 2 |
| Tidriri et al. (2021) | – | 42 000 | 0 | 1 | 2 |

[a] The failure in T11 appears 11 days after a start of the `test2` period. Therefore the maximal possible savings measured within `test2` by random maintenance cannot exceed 3117 €.

1998) because it has only one parameter for an alarm. We restart and re-calibrate DI-CUSUM after known failures to reduce the number of false positives. Fourth, to overcome the CUSUM's dependence on the history of previous alarms, we sample different thresholds with different probabilities. The probabilities are proportional to the profit they give in the training period.

In order to set a scale for comparison between different models, we defined the lower and the upper bounds of possible acceptable maintenance costs. For this we take the natural bounds: the theoretically possible minimum, when all failures are found at a day of maximal savings without any false alarms — `maximal savings`, and the `reactive maintenance` approach, when one does not make any prediction and preventive maintenance, but pays for failures which occurred. To control for random effects, we checked the costs of the `random maintenance`.

Our model allows us to get the results for all turbines and considered periods. We obtain the best savings for both single turbines and the total set of turbines in all periods available for comparison with other authors, except for one case of T07 in the `test2` period. Average costs obtained in our model are better than `random maintenance` in 7 out of 9 cases, and in the remaining 2 cases, both approaches agree within one standard deviation. The uncertainty is significant because of the scarcity of input data and the simplicity of the chosen approach, which contains only one tune parameter. However, our model gives the results that have the mean and even the third quartile of the cost distribution below `reactive maintenance`, which means profitability in most cases. The average savings obtained for the hydraulic group amount to 85 200 Euro for the considered 5 turbines in the test year, with even more significant potential savings of up to 173 100 Euro per year.

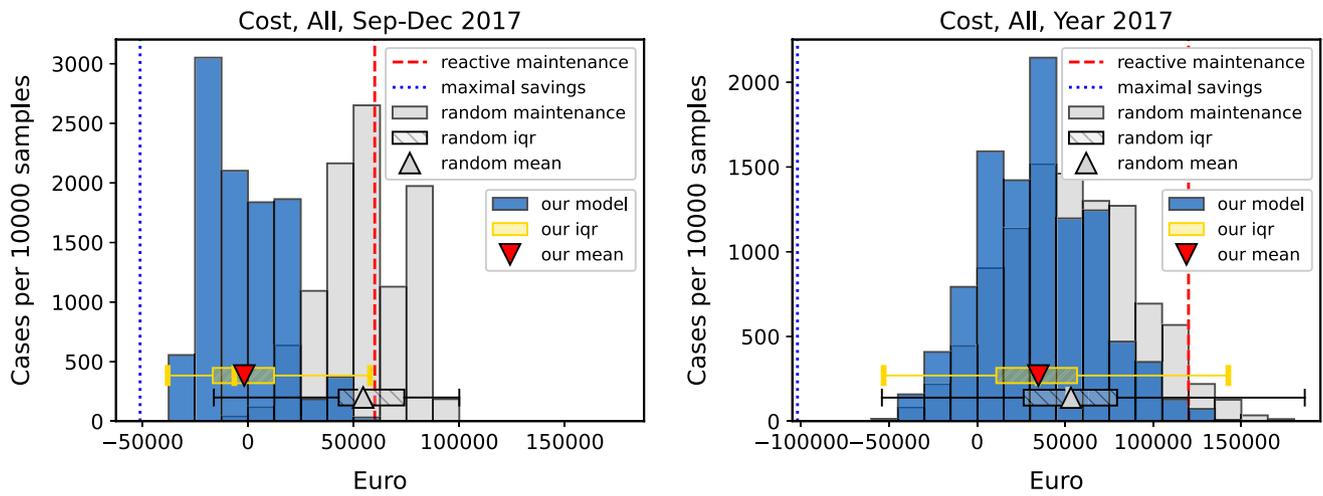

**Fig. 11.** Cost distributions for all five turbines in our model compared to `random maintenance`.

**Table 5**
Total cost for all turbines together.

| Models for all turbines | Cost in Euro | | $\langle \Delta t \rangle$ | $N_{FP}$ | $N_{FN}$ |
|---|---|---|---|---|---|
| | average ± std | min(cost) | | | |
| | | *test1*, Jan–Aug 2017, 3 failures | | | |
| maximal savings | −51 000 | −51 000 | 60 | 0 | 0 |
| random maintenance | 36 400 ± 27 500 | −37 017 | 57.(6) | 6 | 0 |
| our model | 46 200 ± 24 400 | −18 483 | 52.(3) | 13 | 0 |
| reactive maintenance | 60 000 | 60 000 | 0 | 0 | 3 |
| | | *test2*, Sep–Dec 2017, 3 failures | | | |
| maximal savings | −51 000 | −51 000 | 60 | 0 | 0 |
| our model | −1 700 ± 18 000 | −37 950 | 47 | 1 | 0 |
| random maintenance[a] | 54 500 ± 19 700 | −16 116 | 23.(6) | 2 | 0 |
| reactive maintenance | 60 000 | 60 000 | 0 | 0 | 3 |
| | | *test1+2*, year 2017, 6 failures | | | |
| maximal savings | −102 000 | −102 000 | 60 | 0 | 0 |
| random maintenance | 53 300 ± 37 800 | −54 083 | 34.1(6) | 2 | 0 |
| our model | 34 800 ± 32 500 | −53 083 | 44.1(6) | 11 | 0 |
| Eriksson (2020) v0 | – | −21 342 | 40.8(3) | 24 | 0 |
| Eriksson (2020) v1 | – | 85 367 | 9.6(6) | 1 | 0 |
| reactive maintenance | 120 000 | 120 000 | 0 | 0 | 6 |
| Tidriri et al. (2021) | – | 132 000 | 0 | 6 | 6 |

[a] The failures in the `test2` period appear in T07, T09, and T11 only 48, 15, and 11 days after the start of the period, correspondingly. It is less than the earliest allowed detection in 60 days. Therefore, the earliest possible detection limits savings measured within the `test2` period.

We want to emphasize the importance of showing not only the best results, but also the average, standard deviation, and whole distribution of results. Otherwise, one can achieve the best performance randomly after enough sampling, even from a uniform distribution. This approach gives the best minimal costs in all test periods except for the second test period. Also, this exception appears only because of an ambiguity in counting costs in this period. Therefore, it is important to show not only the best possible outcome that a researcher can get for a specific data set with specially fitted hyper-parameters, but also a whole distribution of possibilities. It is relevant because a practitioner will use a model on similar, yet different, real-world data. It seems impossible that a large set of tuned hyper-parameters will coincide for different data sets.

Moreover, researchers rarely publish even the main parameters of their models. On the contrary, by publishing distributions of results, one gives an understanding of a whole range of possible outcomes. Such an approach improves the interpretability and reproducibility of results.

As possible developments in the model, we would like to mention an opportunity to further improve the costs by filtering false alarms. One can straightforwardly use our model for other sensor groups with multiple sensors. This usage would require the calibration of these groups as a whole and comparison with signals from other groups. One could do this using Mahalanobis distance or some other collective distance measure for the whole group together with a decision mechanism that will tell in which sensor group a failure is happening. One can also improve the existing method by considering more sophisticated CUSUM-based approaches, which count a number of steps since a deviation started, moving summation window (MOSUM), or decay of a signal, as in exponentially weighted moving average (EWMA), and similar approaches. However, one should always consider the limited number of failures, which does not allow the introduction of too many new parameters. It would be interesting to see whether the probability of threshold selection based on cost is a good prior for Bayesian algorithms. It would be also interesting to see how well our approach generalizes when more information about failures becomes available. For example, one can take a subset of turbines in a big wind farm and study the uncertainty that appears because of a small set of turbines. The outcome may help develop better maintenance strategies for small wind farms.

An imbalanced dataset is a very challenging but common and important case. Large and expensive systems with many components are made to be reliable. However, even they fail sometimes and produce precisely this type of data – many points in normal behavior and only a few failures. The developed method may be helpful for other expensive mechanisms with multiple sensors, such as maritime and train engines, industrial robots, automated assembly and production lines, and other reliable machines.

**CRediT authorship contribution statement**

**Viktor Begun:** Writing – review & editing, Writing – original draft, Visualization, Validation, Software, Methodology, Investigation, Formal analysis, Data curation, Conceptualization. **Ulrich Schlickewei:** Writing – review & editing, Supervision, Resources, Project administration, Methodology, Funding acquisition, Conceptualization.

**Declaration of competing interest**

The authors declare that they have no known competing financial interests or personal relationships that could have appeared to influence the work reported in this paper.

**Acknowledgments**

We thank Sergiy Begun and Grygorii Shurenkov for fruitful discussions. This research was funded by the Bavarian State Ministry of Science and the Arts (Bayerisches Staatsministerium für Wissenschaft und Kunst), Germany.

**Appendix A. Accumulated random walk in one dimension**

In this appendix, we clarify the distinction between the typical deviation of a single measurement and much larger deviations that can result from cumulative random walks, see Qiu (2013) for more details. The values of CUSUM, which a random walk may reach, depend on an underlying distribution, which leads to that random walk. The simplest example of such a distribution is a Gaussian random walk – a symmetric random walk with a step size $x$, which varies according to a normal distribution with zero mean and standard deviation $\sigma$:

$$P(x) = \mathcal{N}(0, \sigma^2) = \frac{1}{\sigma\sqrt{2\pi}} \exp\left(-\frac{x}{2\sigma^2}\right) . \tag{30}$$

Then, after $n$ steps, the distribution of the traveled distance, $C_n = \sum_{i=1}^{n} x_i$, the CUSUM, will be

$$P(C_n) = \mathcal{N}(0, n\sigma^2) . \tag{31}$$

One can then find an average CUSUM as a root mean squared distance traveled by a Gaussian random walk:

$$\langle C_n \rangle = \sqrt{Var(C_n)} = \sigma\sqrt{n} . \tag{32}$$

It means that, on average, a deviation as large as $\sigma\sqrt{n}$ can appear randomly. One can also estimate a probable upper limit of a CUSUM using the law of the iterated logarithm (Khintchine, 1924; Kolmogoroff, 1929). The magnitude of oscillations in a random walk, governed by a distribution with not too heavy tails and a zero drift, is (Borovkov and Borovkov, 2008):

$$\limsup_{n \to \infty} |C_n| = \sqrt{2\log(\log(n))} * \sigma\sqrt{n} . \tag{33}$$

The $\log(\log(n))$ is a very slowly growing function. If a random walk continues in a range of 1 to 100 days, then the maximal number of steps with 10-minute intervals in one direction is up to 144 and 14 400, correspondingly. For such numbers, the square root with the double logarithm in (33) is approximately constant and is equal to 2. Therefore, one can estimate a maximal CUSUM, which may appear randomly as

$$C_n \sim 2 * \sigma\sqrt{n} . \tag{34}$$

The same applies if fluctuations are not around zero but around $\mu_0$. Reaching a threshold $C_n = h$ can indicate a shift from the starting mean $\mu_0$ to a new value $\mu_1$. To be sure that it is not random, we require that the reached value divided by the number of steps is larger than that estimated in Eq. (34):

$$\mu_1 = \mu_0 + \frac{h}{n} \geq \mu_0 + 2\frac{\sigma}{\sqrt{n}} . \tag{35}$$

The law of the iterated logarithm does not forbid larger deviations but says they are rare (Brown and Petrov, 1975). Therefore, one can replace the coefficient 2 in Eq. (35) with an arbitrary coefficient $\kappa$. The starting assumption was that the distribution is symmetric, so we can rewrite Eq. (35) as follows:

$$|\mu_1 - \mu_0| \geq \kappa \frac{\sigma}{\sqrt{n}} , \tag{36}$$

where one may recognize a Shewhart control chart (Basseville and Nikiforov, 1993). Thus, a new mean, $\mu_1$, should be accepted if it is $\kappa$ times larger than the standard deviation $\sigma$ over the square root of the number of steps $n$. Note that the non-scaled threshold $h$ can be very large. If we take 1 and 100 days as a reference again, then according to Eq. (32), the *average* CUSUMs obtained randomly may reach $\sqrt{144} * \sigma = 12 * \sigma$, and $\sqrt{14\,400} * \sigma = 120 * \sigma$ correspondingly. Moreover, our model gives random fluctuations on the scale of $\sigma = 3$ °C, see Eq. (12), and the average profitable threshold of an accumulated deviation $\langle h \rangle = 19\,400$ °C $\simeq 6\,400 * \sigma$. These numbers differ dramatically from the usual 2-, 3-, and 6-sigma deviations taken for single, non-accumulated deviations. Therefore, the dependence of a CUSUM on the number of steps must be considered.

**Appendix B. DI-CUSUM**

In this appendix, we make a short derivation of the DI-CUSUM formula, see also (Qiu, 2013). If a threshold $h$ is reached, then it is a consequence of a continuous rise or fall (run) of a CUSUM until the value of that threshold $h$. The start of this run can be determined as the last crossing with zero, $i_0$, and Eq. (18) can be rewritten:

$$C_a = \sum_{i=1}^{i_0} (e(t_i) - \mu_0) + \sum_{i=i_0}^{i_a} (e(t_i) - \mu_0) = \sum_{i=i_0}^{i_a} (e(t_i) - \mu_0) = h . \tag{37}$$

The $i_0$ is special because the CUSUM before $i_0$ in Eq. (37) disappears since it is zero at $i_0$ by the definition of $i_0$. Changes that lead to an alarm are assumed irreversible. Therefore, a CUSUM needs to be restarted after reaching a threshold. A new CUSUM should start from a new mean, $\mu_{new} = \mu_{old} + \delta$, to which the residuals have drifted during the time before the alarm. The simplest way to get a new mean is to assume that the change in the CUSUM of the residuals was linear, adding a small constant value at each step. Then,

$$\mu_{new} - \mu_{old} = \delta = h/n , \quad n = i_a - i_0 , \quad \text{and}$$

$$\mu_{new} = \mu_{old} + h/(i_a - i_0) , \tag{38}$$

where $h$ in Eq. (38) comes with its sign: if CUSUM was growing before the alarm, then $h$ is positive; if CUSUM was falling, then $h$ is negative.

**Data availability**

The data are available following the link in the paper.